**Fine-tuning a vision-language model for fracture-surface morphology recognition**

Quanliang Liu[1], Jungtaek Kim[2], Kangwook Lee[2,3,4], Hyunseok Oh[1]*

[1]Department of Materials Science & Engineering, University of Wisconsin–Madison, Madison, WI, USA, 53706

[2]Department of Electrical & Computer Engineering, University of Wisconsin–Madison, Madison, WI, USA, 53706

[3]KRAFTON, Gangnam-gu, Seoul, Republic of Korea, 06142

[4]Ludo Robotics, San Francisco, CA, USA

*Corresponding author. Hyunseok Oh (hyunseok.oh@wisc.edu)

## Abstract

Vision-language models (VLMs) have shown strong potential for scientific image understanding, but general-purpose models often lack the domain-specific visual knowledge required for reliable materials characterization. In this work, we fine-tuned an open-source VLM (Qwen3-VL-32B-Instruct) for fracture-surface image analysis using a curated dataset of 13,168 open-source, literature-mined fracture-surface images. Morphology annotations were generated by GPT-5.2-Reasoning (high) from both the images and relevant excerpts of their source papers, and the dataset was further enriched with targeted manual collection and rotation-based augmentation. The resulting specialist model outperforms flagship proprietary multimodal models on a benchmark of 100 manually annotated images. It achieves a precision of 0.92, compared to 0.35 for the base Qwen3-VL-32B-Instruct, 0.58 for GPT-5.5-Reasoning (high), and 0.78 for Gemini 3.1 Pro-Reasoning (high). Dataset ablations show that manual collection of rare-feature images and augmentation via image rotation are both beneficial to improve recognition of less common fracture morphology features. We further discuss integrated use of the fine-tuned model with proprietary models to combine fracture-specific visual accuracy with broader multimodal reasoning for autonomous fractography. Although focused on fracture-surface images, this work demonstrates how VLMs can be adapted through targeted collection and fine-tuning on novel feature images to recognize those features and support downstream decision-making in autonomous microscopy workflows.

## 1. Introduction

Autonomous microscopy has advanced rapidly through the integration of imaging [1], analysis [2], and decision-making [3]. DeepSPM [4] utilizes convolutional neural networks (CNNs) to classify the probe state and deep reinforcement learning to autonomously execute conditioning protocols during low-temperature scanning tunneling microscopy (STM) imaging of MgPc molecules on Ag(100), enabling stable operation without human intervention. Bayesian active learning subsequently provided a more strategic decision rule for piezoresponse force microscopy/spectroscopy (PFM/S) measurements of $PbTiO_3$ thin films: given uncertainty, prior knowledge, and a scientific objective, choose where to measure next [5]. This optimization-based logic has since been extended from where to measure to how to measure, by defining a reward function that encodes human-operator heuristics to automate tapping-mode scanning probe microscopy (SPM) parameters (e.g., drive amplitude and setpoint) tuning across diverse samples, including $TiO_2$ nanoparticles and $CaCl_2$-containing water droplets on mica [6]. Across these examples, autonomous microscopy has largely relied on narrow perception models paired with hand-crafted objective functions, leaving open how richer, language-grounded interpretation might fit into the loop.

Recently, complementary to these narrow perception models, vision-language models (VLMs) have emerged from a different research thread in general-purpose AI. CLIP [7] shows that through contrastive training on image-text pairs, the model can learn transferable visual representations. Flamingo [8] and BLIP-2 [9] connect strong pretrained vision encoders to frozen large language models (LLMs) via lightweight adapters. This bridging enables Flamingo to perform strong few-shot in-context learning without per-task fine-tuning, while BLIP-2 achieved competitive zero-shot image-to-text and VQA performance. LLaVA [10] then shows that visual instruction tuning could extend this architecture into a general-purpose multimodal assistant by training on GPT-4-generated image-instruction data.

VLMs are thus beginning to take an active role in autonomous experimentation. In CRESt [11], a VLM monitors process cameras and SEM imaging to generate hypotheses that diagnose and correct experimental anomalies, closing the loop with robotic synthesis and knowledge-assisted Bayesian optimization [12] to screen more than 900 multielement electrocatalyst chemistries. EAA [13] expands the VLM's role across the synchrotron x-ray fluorescence (XRF) microscopy pipeline: the model takes the user's command, interprets live images, and issues instrument commands directly for tasks such as zone-plate focusing on Cr-patterned silicon nitride targets and natural-language feature search.

These studies show the importance of an accurate image-understanding module. A wrong interpretation from a VLM can lead to an incorrect instrument command, a flawed hypothesis, or a wasted experiment. What these systems need is a domain-specialized perception module that can supply accurate interpretation while leaving broader reasoning to a generalist model. Yet general VLMs have limitations in specific domain tasks. In materials science, MaCBench [14] shows that strong multimodal models still struggle with spatial reasoning, cross-modal synthesis, and multi-step inference in chemistry and materials tasks. MatCha [15] reaches the same conclusion for materials characterization images that performance drops when questions require scientific interpretation rather than shallow understanding.

For domain adaptation of VLMs, biomedicine provides many examples because its literature contains large numbers of images paired with expert descriptions. BiomedCLIP [16] illustrates this approach, pairing a PubMedBERT text encoder with a vision transformer in a contrastive framework to yield representations that transfer across retrieval, classification, and visual question answering. LLaVA-Med [17] builds on this by using BiomedCLIP as its vision encoder and fine-tuning on GPT-4-generated instruction-following data, extending the model from representation learning to open-ended conversational reasoning over biomedical images.

Domain adaptation of VLMs has also been explored for materials microscopy. SEM-VLM [18] uses CLIP-style contrastive training on literature-mined nanomaterial scanning electron microscopy (SEM) image-text pairs for cross-modal retrieval, classification, and interpretable localization of nanoscale features. MicroscopyGPT [19] fine-tunes a vision-language model to describe atomic structures from simulated scanning transmission electron microscopy (STEM) images. These studies demonstrate the promise of adapting VLMs to microscopy.

In this work, we study supervised fine-tuning (SFT) to adapt an open-source VLM for fracture-surface analysis of metallic materials. Fractography provides a practically important and technically tractable testbed for VLMs in autonomous characterization. First, fracture-surface SEM images of metallic materials require recognition of coexisting fracture morphologies, such as dimples, cleavage facets, river patterns, and striations, which can appear in complex and ambiguous forms. This setting therefore calls for fine-grained, multi-label recognition of subtle and variably expressed fracture features rather than coarser image-level classification or idealized structure description. Second, fractography's visual grammar is compact yet diagnostically rich: features such as dimples, cleavage facets, river patterns, and striations encode failure mechanisms, material responses, and loading conditions [20]. Third, fractography is also well suited to literature-derived image-text curation, because fracture-surface captions commonly contain explicit fracture-related terminology. We chose SFT because our curated fracture-surface dataset provides paired images, morphology descriptions, and feature labels that can directly supervise domain adaptation.

SFT turns a general-purpose VLM into a fractography specialist that uses fracture-specific terminology and produces more accurate and consistent outputs. We then use this specialist as a copilot for proprietary multimodal models: the specialist supplies accurate fracture-surface interpretation, while the proprietary model supplies broader reasoning and interaction. For brevity, the fine-tuned Qwen3-VL-32B-Instruct model is referred to as FT-Qwen hereafter. Our work therefore (1) builds a literature-derived fractography image-text dataset, (2) fine-tunes an open-source VLM that matches or outperforms flagship proprietary multimodal models in recognizing and describing fracture morphology, and (3) shows that FT-Qwen can improve the standalone performance of proprietary multimodal models on fracture-surface interpretation. The code, fine-tuned model weights, dataset metadata, and complete benchmarking / decoding-sweep outputs are deposited on Figshare [21], and the prompts are provided in **Supplementary Information** (**SI**).

## 2. Methodology

### 2.1. Dataset construction

The dataset was constructed through the multi-stage workflow summarized in **Fig. 1a**.

#### 2.1.1. Fracture feature vocabulary definition

Before literature collection, we defined a vocabulary consisting of 11 fracture-surface features: dimple, cleavage, river pattern, intergranular facet, shear lip, striation, beach mark, ratchet mark, fish-eye, chevron, and inclusion particle. Representative examples of each feature are shown in **Fig. 1b**. Labeling allowed multiple features per image, since multiple fracture features can coexist on a single fracture surface. Definitions of the 11 features are provided in **Note S1** of the **SI**.

#### 2.1.2. Paper collection

As illustrated in **Fig. 1a**, the first step in dataset construction was paper collection. We collected papers using the query "fracture surface AND scanning electron NOT polymer NOT ceramic NOT rock" from Elsevier, Springer Nature, AAAS, Wiley, and Taylor & Francis. The goal was to capture metallic SEM fractography while excluding adjacent domains with different fracture morphologies, such as polymers, ceramics, and rocks. The collected papers included materials distributed under multiple open-access licenses, including CC0, CC BY, CC BY-ND, CC BY-NC, CC BY-SA, CC BY-NC-SA, and CC BY-NC-ND; paper metadata were recorded during dataset construction. After an initial round of figure extraction and annotation revealed a long-tail distribution of fracture features, we conducted a second, targeted literature search for papers under the same licenses featuring underrepresented morphologies. From these papers, we manually collected high-quality single-panel images containing rare fracture features: beach mark (137), ratchet mark (94), fish-eye (112), striation (91), inclusion particle (76), chevron (41), and river pattern (28). These values are pre-split targeted-collection counts. After the 100-image hold-out test split, only the training-set portion of these targeted examples is reported in the **Extra collection** in **Fig. 2a**. This targeted collection step was intended to ensure that the dataset contained enough representative examples of uncommon fracture morphologies.

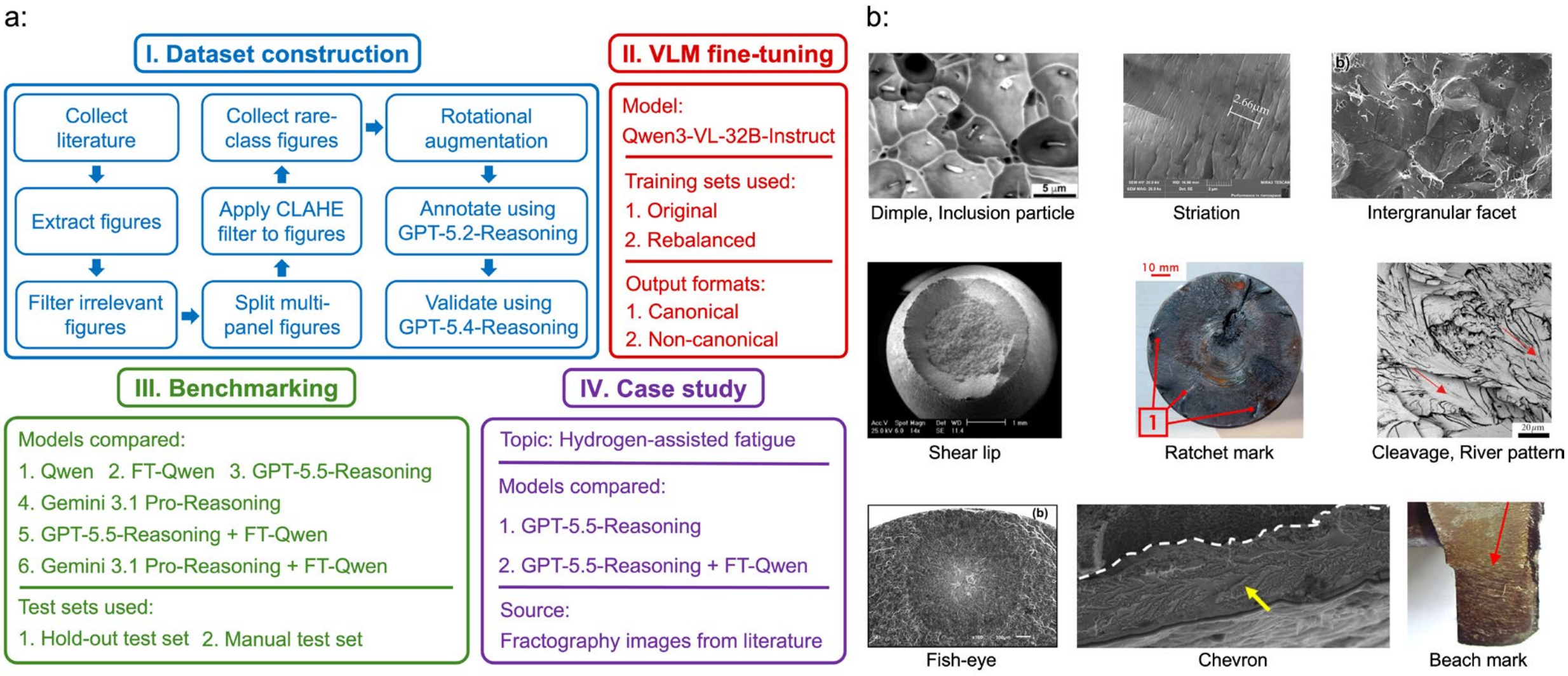

**Figure 1. End-to-end workflow and fracture-feature vocabulary for VLM-based fracture-surface analysis**. (a) The workflow consists of four stages. (I) Dataset construction, (II) VLM fine-tuning: Qwen3-VL-32B-Instruct is fine-tuned using the **Original** and **Rebalanced** training sets under **Canonical** and **Non-canonical** output formats, (III) Benchmarking of the base Qwen model, FT-Qwen, GPT-5.5-Reasoning, GPT-5.5-Reasoning + FT-Qwen, Gemini 3.1 Pro-Reasoning, and Gemini 3.1 Pro-Reasoning + FT-Qwen. All models are evaluated on a hold-out test set and a manual test set. Both proprietary reasoning models were run with reasoning set to high. (IV) Case study: GPT-5.5-Reasoning alone is qualitatively compared with GPT-5.5-Reasoning integrated with FT-Qwen on hydrogen-assisted fatigue images. (b) Representative examples of the 11 fracture-surface features that make up the vocabulary, organized into nine sub-panels: (b1) dimple and inclusion particle, (b2) striation, (b3) intergranular facet, (b4) shear lip, (b5) ratchet mark, (b6) cleavage and river pattern, (b7) fish-eye, (b8) chevron, and (b9) beach mark. All inset images are adapted with CC BY permission for Refs. [22–30], for sub-panels b1-b9 respectively.

### 2.1.3. Figure extraction and filtering

We then used MinerU [31] to extract caption-figure pairs from the downloaded PDFs. As many figures in the retrieved papers were unrelated to fractography, we filtered the extracted caption-figure pairs using a Python regex and retained only those whose caption text matched the string “fract”. The retained set was then manually audited to discard residual non-fractographic figures that passed the filter, such as schematics, plots, and unrelated micrographs. Together, these two steps removed most non-fractography figures while preserving a broad candidate set for subsequent panel extraction and annotation.

a

**Training set distribution**

| Class | Original | | | | Rebalanced |
|---|---|---|---|---|---|
| | Initial collection | Extra collection | Rotation augmentation | Total | Total |
| dimple | 7,147 | 3 | 15 | 7,165 | 925 |
| cleavage | 2,339 | 33 | 27 | 2,399 | 925 |
| striation | 138 | 85 | 702 | 925 | 925 |
| beach mark | 56 | 132 | 600 | 788 | 788 |
| ratchet mark | 52 | 88 | 453 | 593 | 593 |
| fish-eye | 21 | 103 | 405 | 529 | 529 |
| inclusion particle | 269 | 72 | 138 | 479 | 479 |
| intergranular facet | 472 | 0 | 3 | 475 | 475 |
| river pattern | 384 | 27 | 0 | 411 | 411 |
| shear lip | 350 | 6 | 30 | 386 | 386 |
| chevron | 28 | 36 | 225 | 289 | 289 |
| **Total images** | **10,269** | **492** | **2,307** | **13,068** | **5,610** |

b

**Representative image**

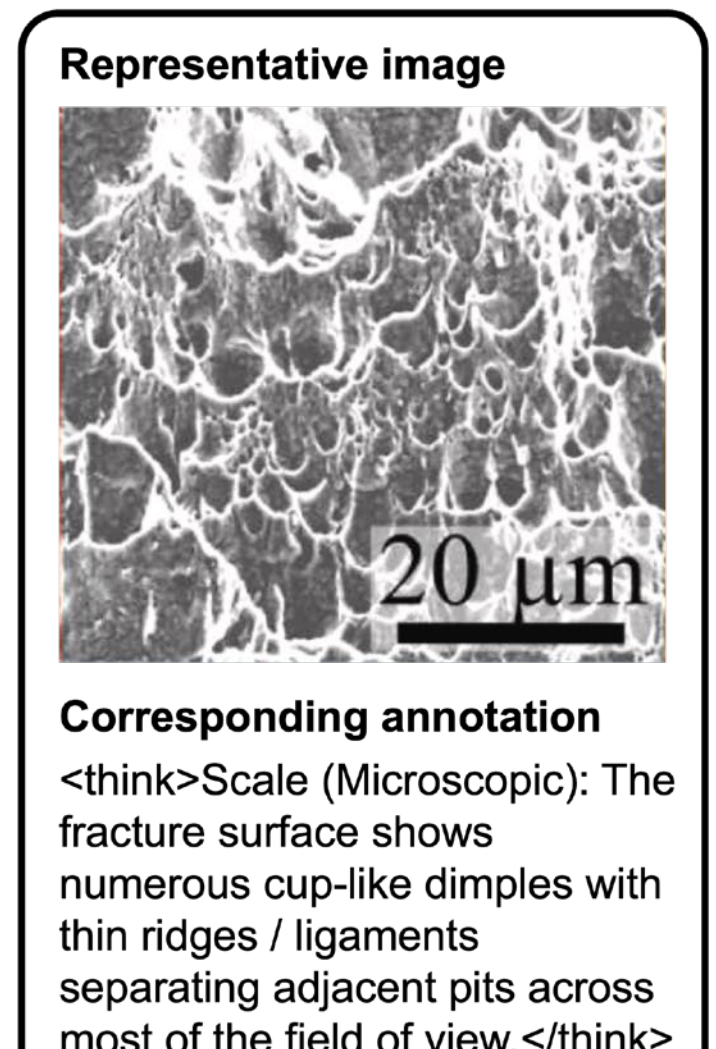

**Corresponding annotation**

<think>Scale (Microscopic): The fracture surface shows numerous cup-like dimples with thin ridges / ligaments separating adjacent pits across most of the field of view.</think>
<answer>dimple</answer>

**Figure 2. Training-set distribution and representative annotated image**. (a) Training-set distribution after the 100-image hold-out split. Under the **Original** training set, the Initial collection, Extra collection, and Rotation augmentation columns report the number of image entries containing each feature from the initial literature-mined set, targeted extra collection, and rotation-generated images, respectively; Total gives their sum for each feature. The **Rebalanced** column reports the corresponding feature counts after downsampling the overrepresented dimple and cleavage classes. The bottom row reports total image entries by source and training-set total. Because labels are multi-label, feature counts are not additive. (b) Representative annotated entry from the dataset, illustrating the multimodal instruction format. Annotation featuring a morphological rationale within <think> tags, specifying scale and visual characteristics, followed by the corresponding label(s) within <answer> tags. The image is adapted with CC BY permission for Ref. [32].

### 2.1.4. Figure pre-processing

Retained compound figures were split into model-friendly single panels because multi-panel layouts can confuse the model and weaken the correspondence between an image and its label. Panel separation was performed using a combination of EXSCLAIM!-assisted splitting [33] and manual splitting, since automated segmentation was not fully reliable for all figure layouts. After panel separation, each single-panel image was contrast-enhanced with contrast-limited adaptive histogram equalization (CLAHE; clip limit = 3, grid size = 8 × 8) [34] to improve the visibility of fracture-surface topography while preserving local contrast.

### 2.1.5. Image augmentation

To further increase the representation of underrepresented fracture features in the dataset, we applied rotation augmentation at 90°, 180°, and 270° to images containing striation, beach mark, ratchet mark, fish-eye, and chevron, effectively quadrupling the number of such images. Rotation was selected because the fracture-surface recognition task in this study is not tied to a fixed image orientation, and the transformation preserves the underlying feature labels while increasing visual diversity. Because labeling was multi-label, a rotation-generated image also retained any co-occurring labels from its source image; consequently, nonzero entries in the **Rotation augmentation** column of **Fig. 2a** can also appear for features that were not directly targeted for rotation augmentation. Because rotation augmentation was applied before the train/test split, the training-set rotation counts may exceed three times the sum of the corresponding **Initial collection** and **Extra collection** counts in **Fig. 2a**. Together, the targeted image collection described in **Section 2.1.2** and rotation augmentation were designed to address feature imbalance in the long-tail features.

#### 2.1.6. Annotation and quality control

Annotation was performed with GPT-5.2-Reasoning [35] (model gpt-5.2-2025-12-11), accessed via the OpenAI Batch API from a Python 3.11.14 environment using the official openai Python client library, with reasoning.effort set to high. For each single-panel image, the model was provided with the image itself, a relevant excerpt from the source paper identified using the corresponding figure and panel identifier, and a structured annotation prompt. The annotation output was restricted to two components: first, a single paragraph describing only the fracture morphologies visible in the image, including scale designation as macroscopic or microscopic; second, a new line listing the detected features as comma-separated labels using the exact names from the predefined 11-feature vocabulary. For the manually collected images, the ground-truth features were provided manually by the authors, and GPT-5.2-Reasoning was used only to generate the corresponding morphology descriptions. **Fig. 2b** shows a representative annotated image from the dataset.

We then used GPT-5.4-Reasoning [36] (model gpt-5.4-2026-03-05) as a second-stage filter, accessed through the same Batch API setup as GPT-5.2-Reasoning, to remove images whose fracture features were too ambiguous to support reliable training. This quality-control step was particularly important for literature-derived images, where individual panels may contain only partial or weakly expressed features. All annotation and filtering prompts are provided in **Note S2** of the **SI**.

The GPT-5.2-Reasoning annotation and GPT-5.4-Reasoning quality control processes cost USD 50 and USD 100, respectively; however, the latter figure should be regarded as an estimate rather than an audited total.

#### 2.1.7. Dataset statistics and split

After construction, the **Original** dataset comprised 13,168 image entries across the 11-feature vocabulary. From this dataset, we randomly selected 100 images with a balanced label distribution to serve as a hold-out test set; the remaining 13,068 images formed the **Original** training set summarized in **Fig. 2a**. This training set contained 10,269 initially collected entries, 492 targeted extra-collection entries, and 2,307 rotation-generated entries. In **Fig. 2a**, feature rows give the number of training-set image entries containing each feature in each source category, whereas the bottom row gives the total number of image entries by source. To address class imbalance, we created a 5,610-image Rebalanced training set by downsampling the overrepresented dimple and cleavage classes; the final column in **Fig. 2a** reports the corresponding feature counts in this **Rebalanced** set.

Evaluation was performed using both the hold-out test set and a separate manual test set. We used two test sets because the hold-out set was drawn from the same literature-derived dataset that was annotated and filtered through a VLM-assisted curation pipeline. Since the benchmarked systems are themselves VLM-based, performance on that hold-out split primarily measures consistency with the original curation pipeline. The separate manual test set was therefore introduced as an external check on whether model predictions also remain aligned with human manual judgment. The manual set comprised 100 additional images absent from the original dataset; these images were collected by the authors to ensure a balanced feature distribution and were manually labeled. The detailed feature distributions for both test sets are provided in **Table S1** of the **SI**.

### 2.2. Model fine-tuning

Given the fine-grained and multi-label nature of fractography, we tested two target-output schemas during fine-tuning, as shown in **Fig. 3**. The first was a **Non-canonical** schema, which allowed the model to emit a short morphological rationale followed by a free-form list of predicted features. The second was a **Canonical** schema, which forced the model to output all 11 features explicitly in JSON form.

During benchmarking, only the content of the <answer> field was used for label scoring; the <think> field was retained as a compact visual rationale to encourage morphology-grounded responses during training. Combining two dataset variants, **Original** and **Rebalanced**, with two output schemas, **Non-canonical** and **Canonical**, yielded four fine-tuning configurations.

In pilot experiments, Qwen3-VL-32B-Instruct [37] outperformed Llama-3.2-11B-Vision [38], GLM-4.6V [39], and InternVL3.5 [40], and was therefore selected for final fine-tuning. We intentionally chose an open-source model over a proprietary API because the checkpoint can be downloaded, locally saved, and re-evaluated later without risk of silent model updates or service retirement. This choice improves reproducibility and long-term reliability.

Fine-tuning was performed using the LLaMA-Factory framework [41] with low-rank adaptation (LoRA) [42] on four NVIDIA H100 GPUs on the Purdue Anvil cluster [43]. Across all fine-tuning configurations, both training and evaluation losses decreased rapidly at early stages and then gradually stabilized, indicating stable convergence. The **Canonical** cases generally exhibited lower loss values than the **Non-canonical** cases, consistent with the more constrained JSON output space. Prompts and training settings are provided in **Notes S2 and S3**, respectively; full loss curves are shown in **Figures S1–S4** of the **SI**.

Non-canonical output format:

<think>Scale (Macroscopic/Microscopic): morphology description</think>
<answer> feature 1, feature 2, ... </answer>

Canonical output format:

<think>Scale (Macroscopic/Microscopic): morphology description</think>
<answer> {“dimple”:0, “cleavage”:0, “river pattern”:0, “intergranular facet”:0, “shear lip”:0, “striation”:0, “beach mark”:0, “ratchet mark”:0, “fish-eye”:0, “chevron”:0, “inclusion particle”:0} </answer>

**Figure 3. Output formats used during VLM fine-tuning.** Both schemas share a <think> block that contains the morphological rationale; they differ only in the structure of the <answer> block. In the **Canonical** format, each of the 11 features is assigned a binary value (1 if present in the image, 0 if absent); the example above shows the template with all features set to 0.

### 2.3. Evaluation protocol

All models were evaluated as multi-label classifiers over the 11-feature vocabulary. For each image, the feature prediction part of the model’s response was parsed into a set of predicted labels and compared with the manually assigned ground-truth labels. Precision, recall, and F1 score were computed at the image level and then averaged over the test set. Feature-wise precision and recall were also calculated for each fracture feature to diagnose class-specific strengths and weaknesses.

The base Qwen3-VL-32B-Instruct model and FT-Qwen variants were evaluated using greedy decoding unless otherwise specified. All proprietary-model evaluations were run from a Python 3.11.14 environment using the official openai and google-genai Python client libraries. GPT-5.5-Reasoning [44] was accessed through the OpenAI Responses API (model gpt-5.5-2026-04-23) with reasoning.effort set to high, and Gemini 3.1 Pro-Reasoning [45] was accessed through the Gemini API (model gemini-3.1-pro-preview) with thinking_level set to high. Each request contained one fracture-surface image and one text prompt instructing the model to identify the morphological features present from the predefined 11-feature vocabulary (full prompts in **Note S2** of the **SI**), and the same parsing and scoring pipeline was used for all models.

## 3. Results

**Fig. 4** shows the image-averaged (each image includes multiple features) precision, recall, and F1 score, together with feature-wise precision and recall, on the **manual test set** under the **Non-canonical** output format. For FT-Qwen, the variants shown are trained on the **Original** dataset. Benchmarking results for other experimental conditions are summarized in **Table 1**.

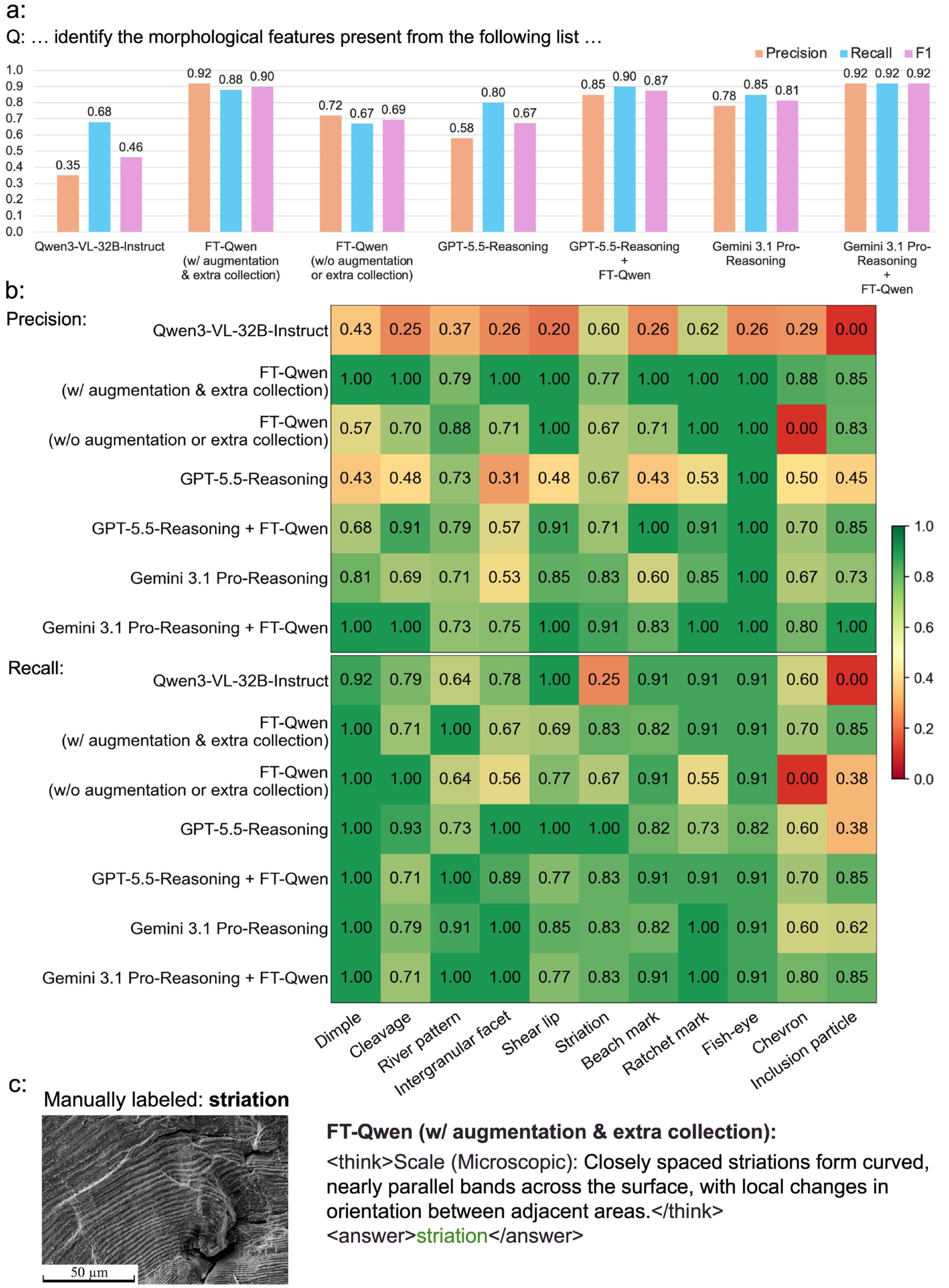

a:

Q: … identify the morphological features present from the following list …

|  | Precision | Recall | F1 |
|---|---|---|---|
| Qwen3-VL-32B-Instruct | 0.35 | 0.68 | 0.46 |
| FT-Qwen (w/ augmentation & extra collection) | 0.92 | 0.88 | 0.90 |
| FT-Qwen (w/o augmentation or extra collection) | 0.72 | 0.67 | 0.69 |
| GPT-5.5-Reasoning | 0.58 | 0.80 | 0.67 |
| GPT-5.5-Reasoning + FT-Qwen | 0.85 | 0.90 | 0.87 |
| Gemini 3.1 Pro-Reasoning | 0.78 | 0.85 | 0.81 |
| Gemini 3.1 Pro-Reasoning + FT-Qwen | 0.92 | 0.92 | 0.92 |

b:

Precision:

|  | Dimple | Cleavage | River pattern | Intergranular facet | Shear lip | Striation | Beach mark | Ratchet mark | Fish-eye | Chevron | Inclusion particle |
|---|---|---|---|---|---|---|---|---|---|---|---|
| Qwen3-VL-32B-Instruct | 0.43 | 0.25 | 0.37 | 0.26 | 0.20 | 0.60 | 0.26 | 0.62 | 0.26 | 0.29 | 0.00 |
| FT-Qwen (w/ augmentation & extra collection) | 1.00 | 1.00 | 0.79 | 1.00 | 1.00 | 0.77 | 1.00 | 1.00 | 1.00 | 0.88 | 0.85 |
| FT-Qwen (w/o augmentation or extra collection) | 0.57 | 0.70 | 0.88 | 0.71 | 1.00 | 0.67 | 0.71 | 1.00 | 1.00 | 0.00 | 0.83 |
| GPT-5.5-Reasoning | 0.43 | 0.48 | 0.73 | 0.31 | 0.48 | 0.67 | 0.43 | 0.53 | 1.00 | 0.50 | 0.45 |
| GPT-5.5-Reasoning + FT-Qwen | 0.68 | 0.91 | 0.79 | 0.57 | 0.91 | 0.71 | 1.00 | 0.91 | 1.00 | 0.70 | 0.85 |
| Gemini 3.1 Pro-Reasoning | 0.81 | 0.69 | 0.71 | 0.53 | 0.85 | 0.83 | 0.60 | 0.85 | 1.00 | 0.67 | 0.73 |
| Gemini 3.1 Pro-Reasoning + FT-Qwen | 1.00 | 1.00 | 0.73 | 0.75 | 1.00 | 0.91 | 0.83 | 1.00 | 1.00 | 0.80 | 1.00 |

Recall:

|  | Dimple | Cleavage | River pattern | Intergranular facet | Shear lip | Striation | Beach mark | Ratchet mark | Fish-eye | Chevron | Inclusion particle |
|---|---|---|---|---|---|---|---|---|---|---|---|
| Qwen3-VL-32B-Instruct | 0.92 | 0.79 | 0.64 | 0.78 | 1.00 | 0.25 | 0.91 | 0.91 | 0.91 | 0.60 | 0.00 |
| FT-Qwen (w/ augmentation & extra collection) | 1.00 | 0.71 | 1.00 | 0.67 | 0.69 | 0.83 | 0.82 | 0.91 | 0.91 | 0.70 | 0.85 |
| FT-Qwen (w/o augmentation or extra collection) | 1.00 | 1.00 | 0.64 | 0.56 | 0.77 | 0.67 | 0.91 | 0.55 | 0.91 | 0.00 | 0.38 |
| GPT-5.5-Reasoning | 1.00 | 0.93 | 0.73 | 1.00 | 1.00 | 1.00 | 0.82 | 0.73 | 0.82 | 0.60 | 0.38 |
| GPT-5.5-Reasoning + FT-Qwen | 1.00 | 0.71 | 1.00 | 0.89 | 0.77 | 0.83 | 0.91 | 0.91 | 0.91 | 0.70 | 0.85 |
| Gemini 3.1 Pro-Reasoning | 1.00 | 0.79 | 0.91 | 1.00 | 0.85 | 0.83 | 0.82 | 1.00 | 0.91 | 0.60 | 0.62 |
| Gemini 3.1 Pro-Reasoning + FT-Qwen | 1.00 | 0.71 | 1.00 | 1.00 | 0.77 | 0.83 | 0.91 | 1.00 | 0.91 | 0.80 | 0.85 |

c:

Manually labeled: **striation**

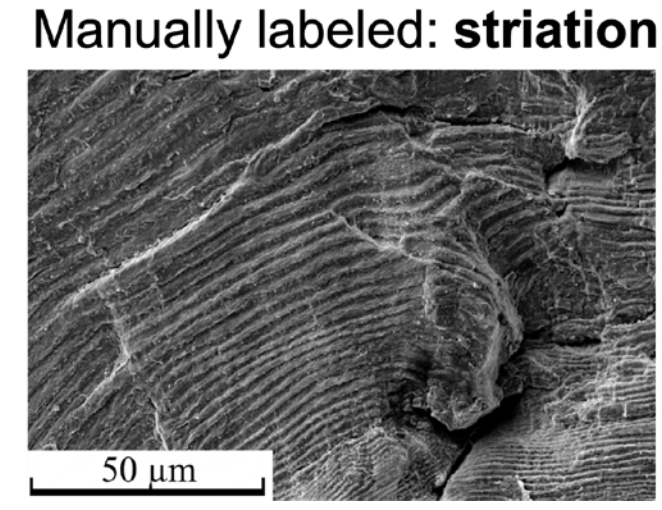

**FT-Qwen (w/ augmentation & extra collection):**

<think>Scale (Microscopic): Closely spaced striations form curved, nearly parallel bands across the surface, with local changes in orientation between adjacent areas.</think>
<answer>striation</answer>

**Figure 4. Benchmarking results on the manual test set. All models follow a Non-canonical output style, and FT-Qwen variants were trained on the Original dataset.** (a) Image-averaged precision, recall, and F1 score for the Qwen3-VL-32B-Instruct, FT-Qwen with augmentation and extra collection, FT-Qwen without augmentation or extra collection, GPT-5.5-Reasoning, GPT-5.5-Reasoning + FT-Qwen, Gemini 3.1 Pro-Reasoning, and Gemini 3.1 Pro-Reasoning + FT-Qwen. Both proprietary reasoning models were run with reasoning set to high. The prompt excerpt shown above the bars indicates the core task given to all models, which is to identify morphological features from the predefined 11-feature vocabulary. The full prompt is provided in **Note S2** of the **SI**. F1 is the harmonic mean of the precision and recall, $\mathbf{F1} = \mathbf{2} * (\boldsymbol{precision} * \boldsymbol{recall})/(\boldsymbol{precision} + \boldsymbol{recall})$, so it summarizes the precision–recall trade-off in a single number. (b) Feature-wise heatmaps of precision (top) and recall (bottom) for the same evaluation conditions across the features. Values are shown in the cells, with greener colors indicating higher performance and redder colors indicating lower performance. (c) Representative FT-Qwen response on a manual test set image labeled as striation, illustrating the structure of a typical model output. The <think> block contains the morphological rationale, and the <answer> block contains the predicted label, here correctly identified as striation (green). The image is adapted with CC BY permission for Ref. [46].

### 3.1. Performance enhancement through domain-specific fine-tuning

The quantitative comparison between the base model and FT-Qwen variants demonstrates the importance of domain-specific adaptation for reliable fractographic analysis (**Fig. 4a**). The base Qwen3-VL-32B-Instruct model showed low precision of 0.35, indicating frequent false-positive predictions. In this context, a false positive occurs when the model incorrectly identifies a fracture feature that is absent from the image. Although the base model had a higher average recall of 0.68, this primarily reflected a tendency to over-predict candidate features rather than robust morphology-specific discrimination.

In contrast, FT-Qwen trained with targeted augmentation and extra image collection achieved a substantial improvement, with average precision and recall reaching 0.92 and 0.88, respectively. This improvement indicates that domain-specific fine-tuning reduced false-positive predictions while maintaining high sensitivity to the target fracture features. The feature-wise heatmap (**Fig. 4b**) further shows that FT-Qwen performed strongly across both common and underrepresented morphologies. For example, the enriched FT-Qwen achieved precision values of 1.00 for dimple, cleavage, intergranular facet, shear lip, beach mark, ratchet mark, and fish-eye, while also maintaining high recall for dimple, river pattern, striation, beach mark, ratchet mark, fish-eye, and inclusion particle. **Fig. 4c** shows a representative FT-Qwen response example on a manual test set image labeled as striation, where the model correctly identifies the feature. The output follows the <think>/<answer> schema introduced during fine-tuning (**Fig. 3**), confirming that fine-tuning successfully aligned the output with the target format.

### 3.2. Performance benchmarking: specialist model versus proprietary generalist models

FT-Qwen trained with dataset augmentation and extra collection achieved 0.92 precision, 0.88 recall, and 0.90 F1 on the manual test set and was benchmarked against two frontier proprietary reasoning models, GPT-5.5-Reasoning and Gemini 3.1 Pro-Reasoning, both evaluated with reasoning set to high (**Fig. 4a**). Although these proprietary models exhibit strong general multimodal reasoning capabilities, both trailed the specialist model in precision, recall, and F1 score. GPT-5.5-Reasoning achieved 0.58 precision, 0.80 recall, and 0.67 F1, indicating that its main limitation was a high false-positive rate. Its feature-wise precision profile shows frequent false-positive predictions for several fracture features, including dimple, cleavage, intergranular facet, shear lip, beach mark, and inclusion particle. Its recall was higher than its precision for many features, reflecting a tendency to over-predict candidate labels rather than reliable feature discrimination.

Gemini 3.1 Pro-Reasoning was the stronger proprietary standalone model, achieving 0.78 precision, 0.85 recall, and 0.81 F1. Compared with FT-Qwen, Gemini achieved comparable recall, indicating similar feature sensitivity. However, FT-Qwen achieved substantially higher precision (0.92), whereas Gemini retained a greater tendency to assign extra fracture-feature labels, especially for intergranular facet and beach mark. Overall, these results show that FT-Qwen provided the best standalone performance among the tested models, achieving the highest precision, recall, and F1 score on the manual test set.

### 3.3. Impact of image augmentation and targeted extra collection

An ablation comparison was conducted to evaluate the contribution of rotation-based augmentation and targeted manual image collection. FT-Qwen trained without augmentation or extra collection reached only 0.72 precision, 0.67 recall, and 0.69 F1, whereas the fully enriched FT-Qwen reached 0.92 precision, 0.88 recall, and 0.90 F1 (**Fig. 4a**). This corresponds to a gain of 0.20 in precision, 0.21 in recall, and 0.21 in F1. Dataset statistics for the training set without augmentation or extra collection are provided in **Table S2** of the **SI**.

The feature-wise results show that the enrichment strategy is especially important for long-tail features. Compared with the model trained without augmentation or extra collection, the enriched model improved precision for beach mark from 0.71 to 1.00 and for chevron from 0.00 to 0.88. It also improved recall for striation from 0.67 to 0.83, chevron from 0.00 to 0.70, and ratchet mark from 0.55 to 0.91. These improvements indicate that the enriched dataset enabled the model to learn more robust visual representations of underrepresented fracture morphologies. Chevron provides a clear example: the model trained without augmentation or extra collection produced no correct chevron detections, whereas the enriched model substantially improved both chevron precision and recall. A few individual feature-wise recall values decreased, such as cleavage recall from 1.00 to 0.71 and beach mark recall from 0.91 to 0.82, but these changes occurred alongside large precision gains, indicating a shift from overly aggressive labeling toward more selective detection. Overall, targeted extra collection and augmentation improved image-averaged precision and recall, indicating fewer false positives and fewer false negatives, improving the model's practical usefulness for fractographic analysis.

### 3.4. Performance improvement of proprietary models through specialist assistance

We examined whether FT-Qwen could serve as an effective assistant to enhance the performance of proprietary generalist reasoning models. In the assisted setting, the output from FT-Qwen was provided to GPT-5.5-Reasoning and Gemini 3.1 Pro-Reasoning as additional context, with the full prompt provided in **Note S2** of the **SI**.

As shown in **Fig. 4a**, this assistance yielded clear performance gains. GPT-5.5-Reasoning improved from 0.58 precision, 0.80 recall, and 0.67 F1 to 0.85 precision, 0.90 recall, and 0.87 F1. Gemini 3.1 Pro-Reasoning also benefited, improving from 0.78 precision, 0.85 recall, and 0.81 F1 to 0.92 precision, 0.92 recall, and 0.92 F1. Thus, the FT-Qwen-assisted setting improved both proprietary models in both precision and recall. For GPT-5.5-Reasoning, assistance raised the F1 score by 0.20, bringing it close to the standalone specialist model. For Gemini 3.1 Pro-Reasoning, assistance raised the F1 score by 0.11, producing the highest overall performance among all tested configurations.

Feature-wise analysis further illustrates the value of specialist assistance. For GPT-5.5-Reasoning, assistance improved precision for several features, including cleavage from 0.48 to 0.91, shear lip from 0.48 to 0.91, beach mark from 0.43 to 1.00, ratchet mark from 0.53 to 0.91, and inclusion particle from 0.45 to 0.85. For Gemini 3.1 Pro-Reasoning, the primary effect of assistance was improved selectivity. Precision increased for dimple from 0.81 to 1.00, cleavage from 0.69 to 1.00, shear lip from 0.85 to 1.00, ratchet mark from 0.85 to 1.00, and inclusion particle from 0.73 to 1.00. These results indicate that FT-Qwen can function as a domain-specific reference signal that helps proprietary models reduce false-positive predictions while even improving recall.

## 4. Discussion

### 4.1. Dataset ablation and the role of data composition

The dataset ablation study shows that the performance of FT-Qwen depends strongly on training set composition. The best performance is obtained when both rotation augmentation and targeted extra collection are included. Removing both components causes the largest decrease in performance, with precision and recall dropping by up to 0.20 and 0.21, respectively, relative to the corresponding fully enriched configuration, as shown in **Fig. 5**. This confirms that the raw literature-mined dataset alone is not sufficient for robust recognition of all fracture features, especially long-tail morphologies.

The relative importance of targeted extra collection and rotation-based augmentation cannot be concluded to a simple ranking, because their effects vary across configurations. For example, in the **Original & Non-canonical** setting, removing extra collection causes slightly larger drops than removing augmentation (0.11/0.11 vs. 0.09/0.09 in precision/recall), whereas in the **Rebalanced & Canonical** setting, removing augmentation causes a slightly larger precision drop than removing extra collection. The **Original & Canonical** setting shows the clearest advantage of extra collection, where removing extra collection decreases precision and recall by 0.08 and 0.11, while removing augmentation has little effect. What is consistent across configurations, however, is that removing either component degrades performance. Thus, both strategies are valuable: augmentation improves robustness to orientation variation, whereas extra collection supplies additional high-quality examples of underrepresented fracture morphologies. The most reliable approach is therefore to use both, as their combination gives the strongest and most stable performance across the tested settings. Dataset statistics supporting these ablation results are detailed in **Table S2**.

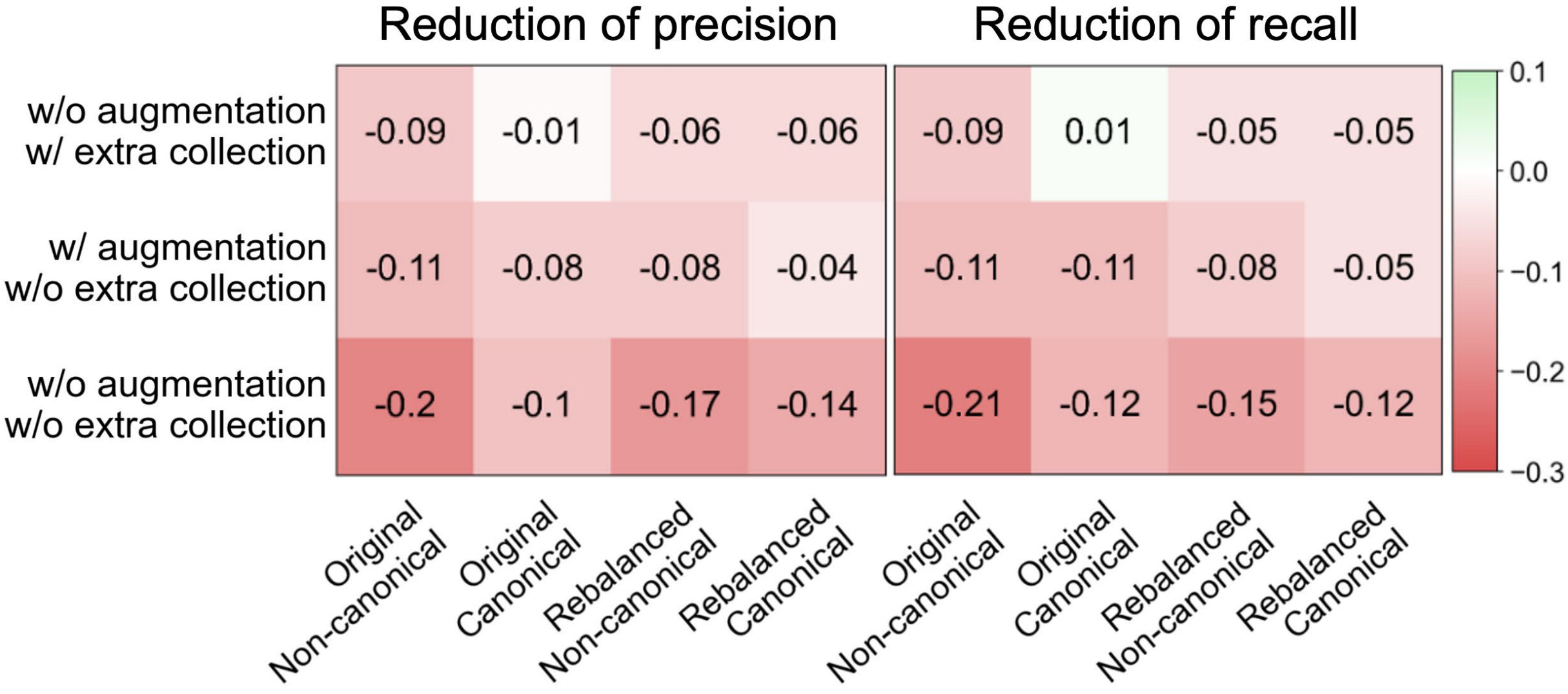

**Figure 5. Dataset ablation study for FT-Qwen.** Per-cell value reductions in image-averaged precision (left) and recall (right) for three fine-tuning variants (w/ augmentation & w/o extra collection; w/o augmentation & w/ extra collection; w/o augmentation & w/o extra collection), each evaluated under the four combinations of dataset split (**Original**, **Rebalanced**) and output schema (**Canonical**, **Non-canonical**). All value reductions are relative to the fully enriched reference configuration (w/ augmentation & w/ extra collection) on the manual test set; more negative (deeper red) entries indicate larger performance loss when that ingredient is removed. The corresponding results for the hold-out test set can be found in **Figure S5** of the **SI**.

### 4.2. Effect of canonical output format and dataset rebalancing

The comparison between **Canonical** and **Non-canonical** output formats shows no consistent advantage for either format; the relative ranking flips depending on the other configuration choices. As shown in **Table 1**, in the fully enriched setting (w/ augmentation and w/ extra collection) on the manual test set, the **Non-canonical** format outperforms the **Canonical** format on the **Original** dataset, with precision/recall of 0.92/0.88 versus 0.85/0.82. However, when augmentation is removed but extra collection is retained on the same **Original** dataset, the trend reverses: the **Canonical** format outperforms the **Non-canonical** format, with precision/recall of 0.84/0.83 versus 0.83/0.79. Similar reversals appear across other configurations and the **Rebalanced** dataset, indicating that neither format is uniformly better when sufficient rare-feature data are available. The one consistent observation is that under heavy data ablation, the **Canonical** output shows smaller performance drops than the **Non-canonical** output. As shown in **Fig. 5**, when both augmentation and extra collection are removed, the precision drop is −0.10 versus −0.20 for the **Original** dataset and −0.14 versus −0.17 for the **Rebalanced** dataset. These results suggest that the explicit 11-label JSON schema may provide regularization under heavy data ablation, but the differences are not large or consistent enough across configurations to conclude that one output format is universally preferable.

The **Original** versus **Rebalanced** dataset comparison similarly shows no uniform effect; the relative ranking again depends on the chosen output format. As shown in **Table 1**, with the fully enriched dataset on the manual test set, under the **Non-canonical** format the **Original** dataset outperforms the **Rebalanced** dataset (precision/recall of 0.92/0.88 versus 0.88/0.85), whereas under the **Canonical** format the trend reverses and the **Rebalanced** dataset slightly outperforms the **Original** (0.86/0.83 versus 0.85/0.82). Rebalancing was introduced to address the dominance of dimple and cleavage in the dataset, but it does not appear to deliver this benefit reliably. Across configurations, the **Original** and **Rebalanced** datasets perform comparably, likely because downsampling removes useful common-feature diversity without adding genuinely new examples of rare morphologies.

### 4.3. Specialist assistance for proprietary models

The model-assistance results in **Fig. 6** show that FT-Qwen can serve as an assistant for proprietary multimodal models in identifying features. When FT-Qwen output is supplied as an additional reference, GPT-5.5-Reasoning improves substantially in both precision and recall. For example, in the **Non-canonical** setting, its precision increases by 0.27 and its recall increases by 0.10. This suggests that the specialist model helps GPT-5.5-Reasoning both suppress false-positive labels and recover fracture features that are otherwise missed.

Gemini 3.1 Pro-Reasoning also benefits from specialist assistance, although the improvement is mainly in precision (+0.14 in **Non-canonical**, +0.11 in **Canonical**) with smaller gains in recall (+0.07 and +0.03). This is consistent with its standalone behavior: Gemini 3.1 Pro-Reasoning already shows high recall but tends to over-predict plausible fracture features. FT-Qwen therefore acts as a selective visual prior, helping the proprietary model avoid unsupported labels while maintaining high sensitivity. These results support a modular workflow in which a fine-tuned specialist model provides fracture-specific perception, while a proprietary reasoning model performs higher-level confirmation, interpretation, and report generation.

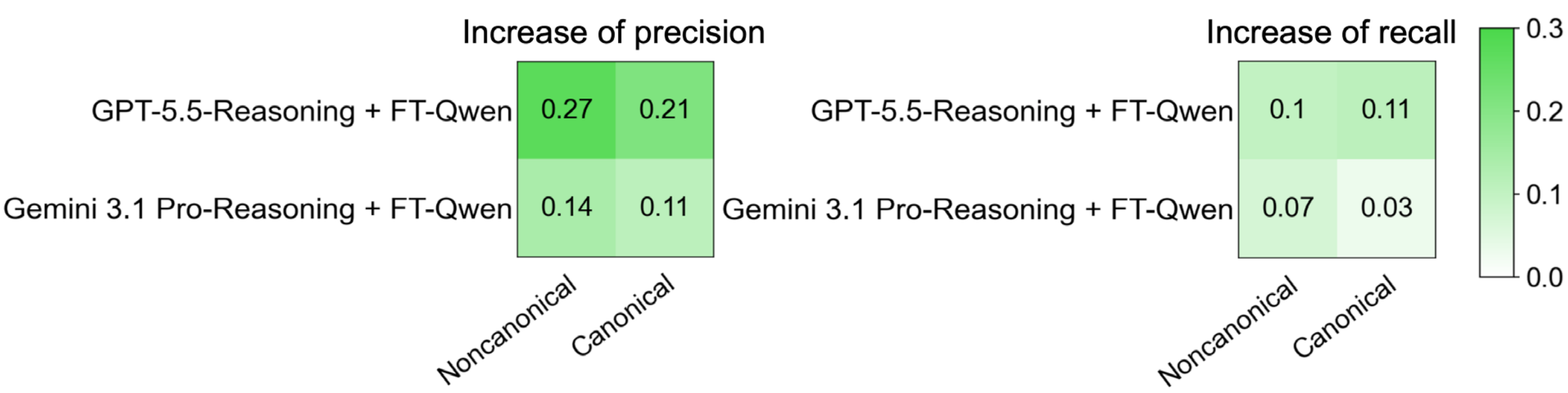

**Figure 6. Specialist-assistance analysis for proprietary models.** Per-cell performance gains in image-averaged precision (left) and recall (right) for GPT-5.5-Reasoning + FT-Qwen and Gemini 3.1 Pro-Reasoning + FT-Qwen relative to their unassisted baselines, each evaluated under two output schemas (**Canonical**, **Non-canonical**) on the manual test set; more positive (deeper green) entries indicate larger improvements from specialist assistance. Both proprietary reasoning models were run with reasoning set to high. The corresponding results for the hold-out test set can be found in **Figure S6** of the **SI**.

### 4.4. Limitations of FT-Qwen

Although fine-tuning substantially improves fracture-feature recognition, FT-Qwen still has limitations. One weakness is the distinction between river pattern and striation, and even between ductile and brittle variants of striation. Striations are cyclic crack-growth markings associated with fatigue [47], whereas river patterns are cleavage-related markings formed by the convergence of cleavage steps [48]. Moreover, striation is itself not a single morphology: ductile striations are produced by alternating slip at a blunting crack tip in relatively inert environments and typically appear as closely spaced, roughly parallel, ridge-like markings, whereas brittle striations are associated with quasi-cleavage fatigue increments promoted by hydrogen and are typically flatter [49]. **Fig. 7** illustrates this challenge: panel (a) shows a river pattern, panel (b) shows ductile striations, and panel (c) shows brittle striations from Refs. [50–52]. Despite their different mechanistic origins, all three can present as lineated markings in cropped SEM images, and the correct assignment often depends not only on subtle morphology but also on contextual cues such as scale, crack-growth direction, surrounding fracture morphology, loading mode, and environment. The ductile-versus-brittle striation distinction therefore cannot always be resolved from image appearance alone, and resolving this ambiguity is beyond the scope of this work.

This limitation is not purely visual-recognition failure but a context-reasoning failure. It may be better addressed by fine-tuning a reasoning-capable multimodal model, because the distinction is not only a matter of visual pattern recognition. Both features can appear as lineated markings, and the correct label often depends on subtle morphology, such as regular periodic spacing in fatigue striations versus converging cleavage steps in river patterns, together with scale, loading mode, crack-growth direction, and environmental condition. A reasoning model can explicitly compare these competing hypotheses, inspect small discriminative features, and verify whether the proposed interpretation is physically consistent with the broader fracture context. This process can be more sample-efficient than relying on image-level pattern learning alone, because contextual fracture-mechanics constraints help narrow the set of plausible labels before the model commits to an answer.

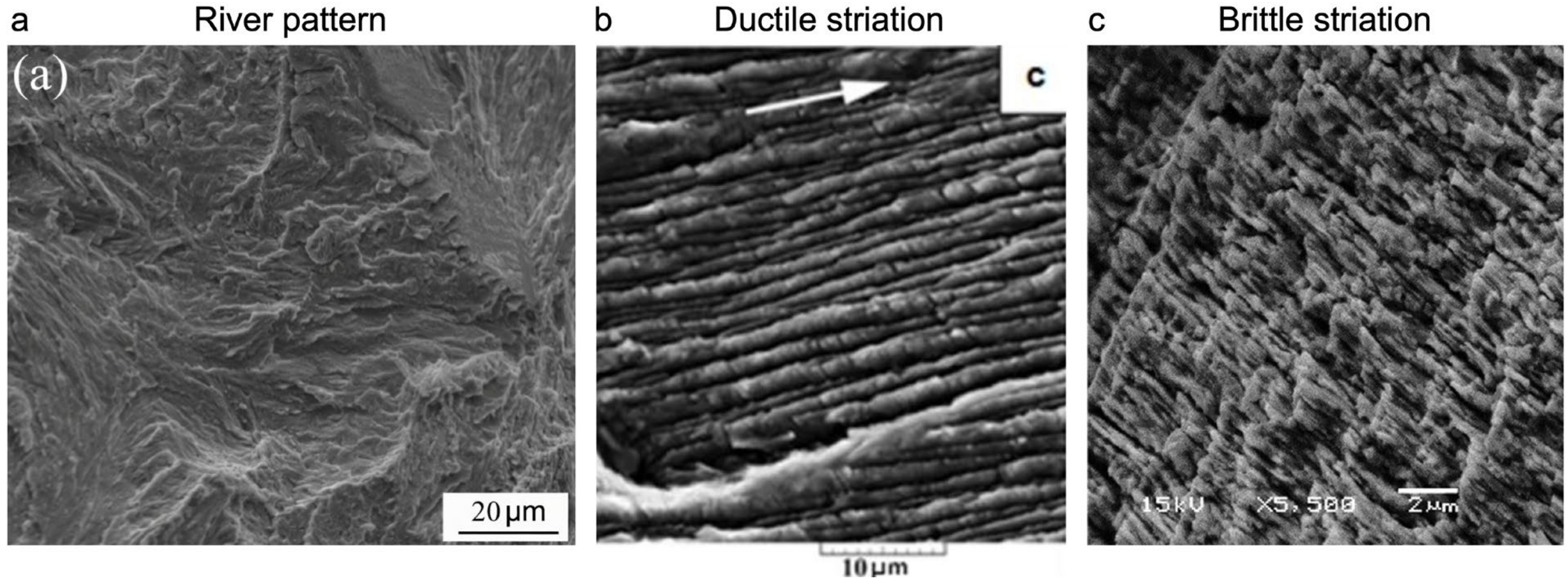

**Figure 7. Visual similarity among lineated fracture-surface morphologies in SEM fractography.** (a) River pattern showing cleavage-step markings that merge, branch, or curve along the local crack-growth direction. (b) Ductile striations showing locally parallel, ridge-like cyclic crack-growth markings. (c) Brittle striations showing lineated fatigue-related morphology with weaker plastic-relief contrast than typical ductile striations. Panels (a)–(c) are adapted with CC BY permission for Refs. **[50–52]**.

### 4.5. Proprietary–specialist integration in fatigue fracture analysis

The qualitative examples in **Fig. 8** further illustrate the value of proprietary-specialist integration in fatigue-fracture analysis: compared with the standalone proprietary model, the integrated workflow produces more mechanistically grounded interpretations of complex mixed-mode fracture surfaces. The **Fig. 8** panel labeled Image 1, obtained under cyclic loading without hydrogen, shows fatigue striations and cyclic crack-growth markings [53]. Standalone GPT-5.5-Reasoning described the surface only in generic terms as a "mixed transgranular tearing / quasi-cleavage fracture with significant plastic deformation", and therefore did not capture the fatigue failure mode. By contrast, the specialist-integrated GPT-5.5-Reasoning response correctly identified "fine, locally parallel striations on transgranular-looking facets" (green text in **Fig. 8**).

The **Fig. 8** panel labeled Image 2, corresponding to the hydrogen condition, shows a more complex mixed-mode fracture appearance. The lower-magnification image acts as a map of transgranular and intergranular features, while the two zoomed-in **Fig. 8** panels labeled Image 2-1 and Image 2-2 reveal distinct striation-like markings on selected grain-boundary surfaces. This morphology is more difficult to interpret because grain-boundary surfaces are often associated with brittle-like intergranular fracture, but the fine markings indicate fatigue-related crack growth. Compared with responses without FT-Qwen, the assisted responses are more specific and mechanistically meaningful, identifying "local striations preserved on grain-boundary facets". The results were also stable across 10 decoding configurations of FT-Qwen (one greedy and nine non-greedy across a temperature / top-p sweep). All prompts used, non-greedy configurations, and the analysis of GPT-5.5-Reasoning outputs are provided in **Note S2**, **Table S3**, and **Figure S7** of the **SI**, respectively.

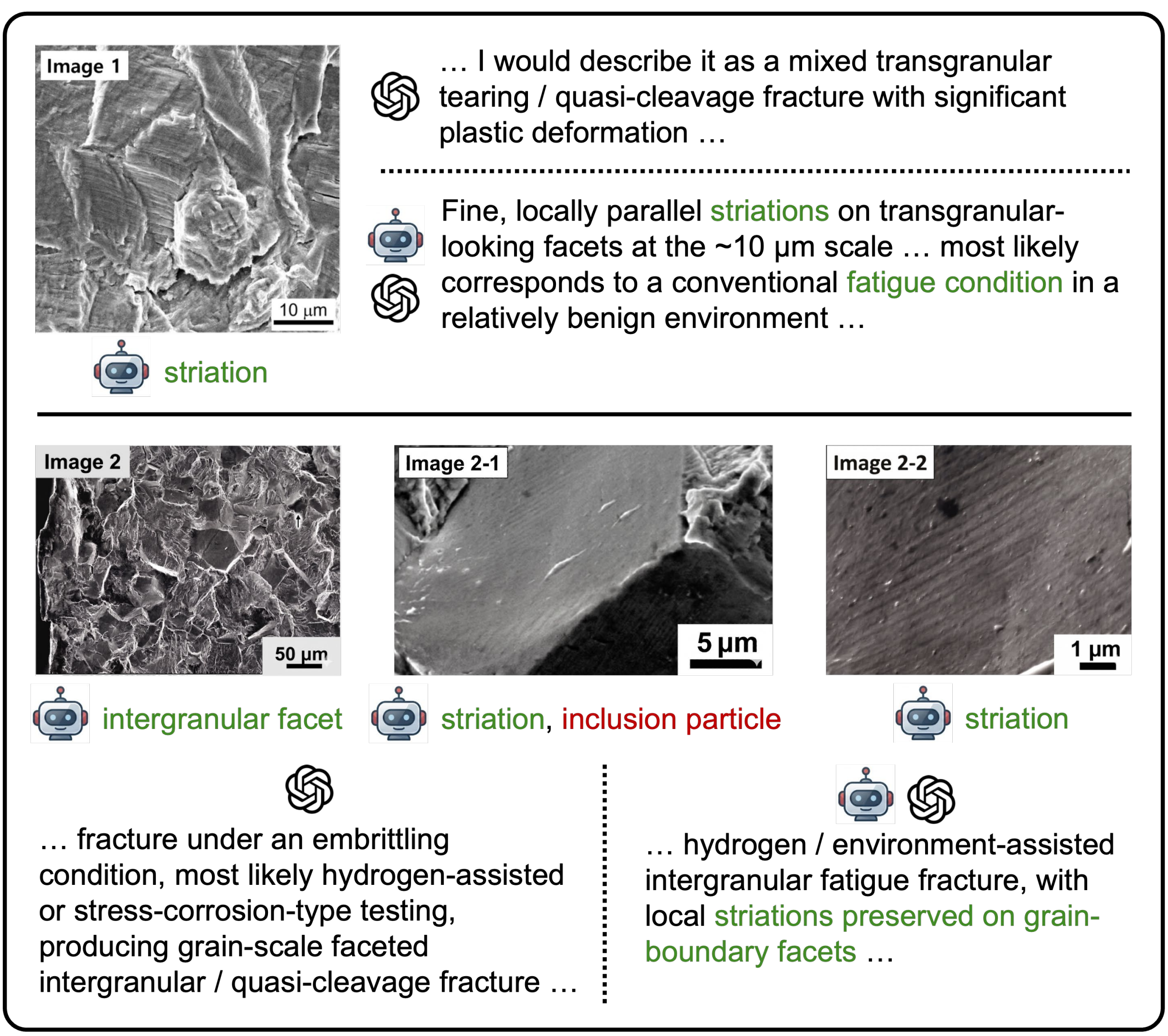

**Figure 8. Qualitative fractographic analysis on an out-of-distribution hydrogen-embrittlement case.** In this figure, the labels Image 1, Image 2, Image 2-1, and Image 2-2 refer to the panels shown here. Images 1, 2, 2-1, and 2-2 are adapted from Ref. [53] with permission from Elsevier. Image 1: fracture surface of the HEA after cyclic loading without hydrogen, dominated by fatigue striations and showing no "brittle-like" intergranular cracking. Image 2: fracture surface after cyclic loading with hydrogen charging, exhibiting a mixed-mode morphology that serves as a macroscopic map of transgranular and intergranular features. Images 2-1 and 2-2: zoomed-in panels on the arrow-indicated region in Image 2, revealing fine, regularly spaced striation-like patterns on the facet surface. Icons next to each output denote the model(s) involved: the robot icon alone marks output produced by standalone FT-Qwen, the GPT icon alone marks output produced by standalone GPT-5.5-Reasoning, and the two icons together mark output produced by GPT-5.5-Reasoning while considering the FT-Qwen prediction for that image. Green text marks correct identifications and red text marks incorrect ones. GPT-5.5-Reasoning was evaluated with reasoning set to high.

### 4.6. Autonomous microscopy blueprint in an agentic AI framework

As shown in **Fig. 9,** the present work can be placed within a broader autonomous microscopy workflow. More broadly, this architecture fits the emerging paradigm of agentic AI, in which language models reason and interact through external tools and environments rather than functioning only as standalone chat interfaces [54,55]. The red-highlighted section represents the contribution of the present study: a fine-tuned VLM specialized in analyzing fracture-surface images. In a future agentic system, SEM images would first be acquired from a fracture surface and then passed to this specialist perception module. Its output could then be passed to a generalist reasoning agent, likely a proprietary model, to combine the visual interpretation with literature-mined processing-structure-property-performance knowledge [56] and mechanistic hypotheses [57] from LLMs. Based on this combined information, the agent could decide where the microscope should image next, for example by zooming in on a suspected region. Dedicated quantitative models could also be integrated to support tasks such as feature segmentation and striation counting [58–61]. From an agent-systems perspective, the specialist model, SEM control API, quantitative analysis tools, and literature-retrieval components together form what recent work terms the agent harness: the external runtime, tool interfaces, and control structure that make the overall system more reliable than a prompt-only workflow [62]. Importantly, this workflow is naturally extensible beyond the current 11-feature vocabulary. When the agent encounters an unknown feature that is not yet represented in the specialist model, the automated SEM can be directed to acquire a large set of images covering the new feature across regions and magnifications. These images can then be processed through the same augmentation pipeline described in **Section 2.1.5** and used to further fine-tune the specialist VLM on an expanded dataset and vocabulary. This closed-loop update allows the perception module to grow alongside the scientific scope of the experiments it supports, so that FT-Qwen increasingly functions as a reliable perception companion to the LLM reasoning agent in fully autonomous SEM workflows.

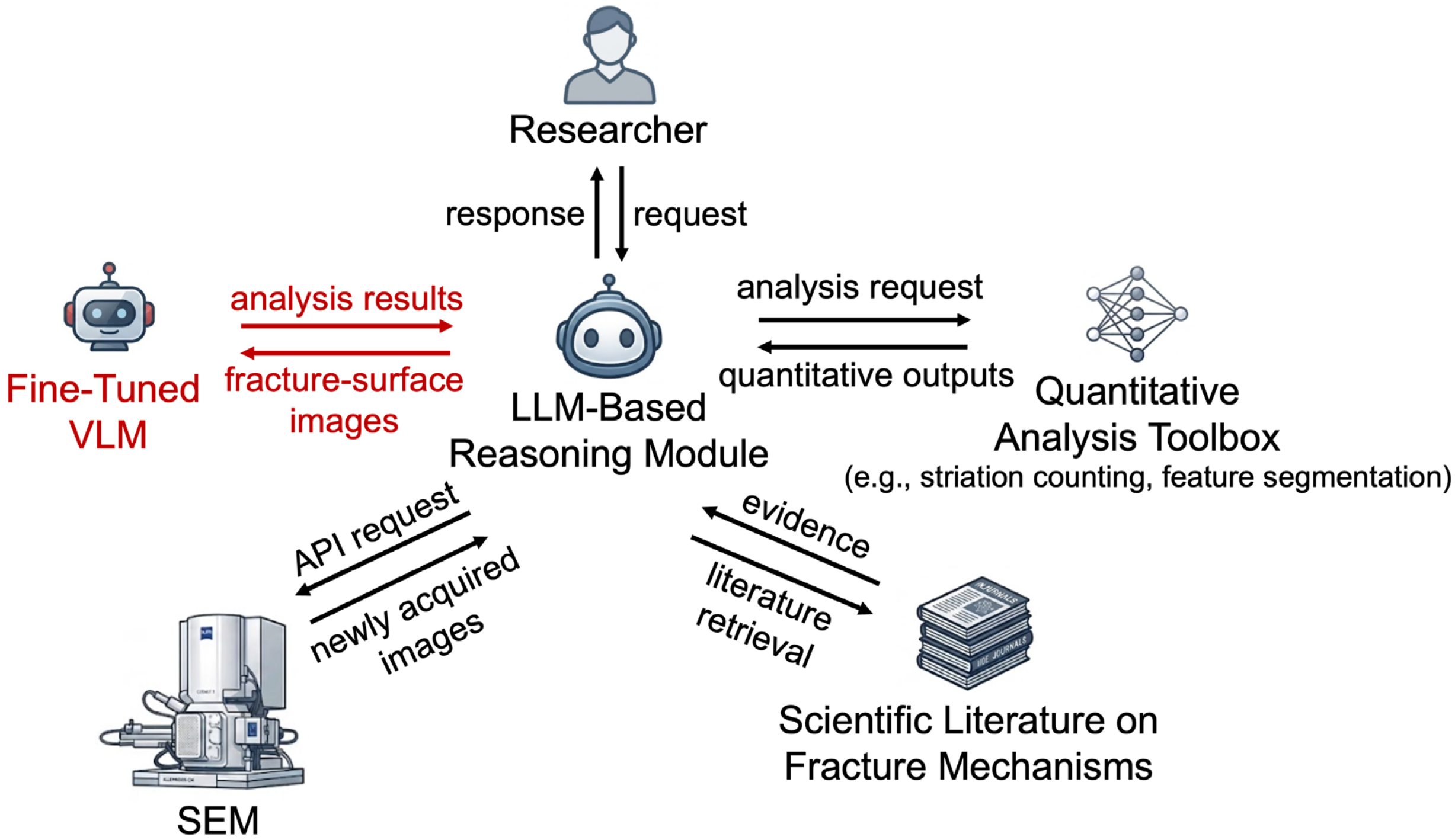

**Figure 9. Blueprint for VLM-in-the-loop autonomous failure analysis.** A central LLM-based reasoning module mediates a four-way interaction between the human researcher, a fine-tuned fractography VLM (red, highlighting the contribution of this work), an SEM accessed through an API for targeted image (re)acquisition, a quantitative analysis toolbox (e.g., striation counting, feature segmentation), and a fracture-mechanism literature base for retrieval-augmented evidence. Arrows indicate the direction of data and control flow between components.

## 5. Conclusions

Overall, this work demonstrates that domain-specific fine-tuning can transform an open-source VLM into a useful specialist for fracture-surface analysis. The following conclusions can be drawn.

First, LoRA-based fine-tuning of Qwen3-VL-32B-Instruct on a literature-mined fractography dataset produced FT-Qwen, which reached 0.92 precision, 0.88 recall, and 0.90 F1 on the 100-image manual test set. FT-Qwen substantially outperformed GPT-5.5-Reasoning (precision 0.58, recall 0.80, F1 0.67) across all three metrics, with the largest gain in precision (+0.34). Against the stronger Gemini 3.1 Pro-Reasoning (precision 0.78, recall 0.85, F1 0.81), FT-Qwen also led on all metrics, again most notably in precision (+0.14). We therefore conclude that FT-Qwen is the stronger standalone model for fracture-surface morphology recognition.

Second, rotation-based augmentation and targeted manual collection of rare-feature images are complementary for recognizing long-tail fracture features, and their combination gives the strongest and most stable performance across the tested settings.

Third, neither output format nor rebalancing strategy is universally preferable: the relative ranking flips across configurations rather than showing a consistent direction, making the choice of output format or rebalancing strategy less impactful than targeted extra collection and augmentation.

Fourth, FT-Qwen also improves proprietary models when supplied as a reference signal: with FT-Qwen assistance, GPT-5.5-Reasoning gained in both precision and recall, while Gemini 3.1 Pro-Reasoning primarily gained in precision. Qualitative analysis of hydrogen-assisted fatigue images further showed that this proprietary-specialist integration produces more mechanistically grounded interpretations than the standalone proprietary model, combining fracture-specific perception with broader multimodal reasoning.

Finally, some distinctions, such as river pattern versus fatigue striation and ductile versus brittle striation, require contextual reasoning beyond visual recognition, and fine-tuning a reasoning-optimized VLM rather than the instruct variant used here is a promising direction for resolving such cases.

Looking further ahead, FT-Qwen, together with quantitative tools and a knowledge base, can be orchestrated by a proprietary model as the central reasoning agent to realize the autonomous fractography workflow, and the same recipe should transfer to general autonomous characterization tasks.

**Acknowledgements**

This work was supported by the National Science Foundation EAGER Award No. 2513480. This work also used resources supported by the National Artificial Intelligence Research Resource under Award No. 250128 and the National Science Foundation ACCESS program under Award No. MAT260014.

**Author contributions**

Q.L. and H.O. conceptualized and designed the study; Q.L. curated the dataset, fine-tuned the vision-language model, developed the code, and conducted the evaluations; J.K., K.L., and H.O. provided technical guidance and support for dataset construction, model fine-tuning, and evaluations; Q.L. and H.O. wrote the paper and the supplementary material; and all authors discussed the results, revised the manuscript, and approved the final version of the manuscript.

**Data availability**

The dataset used in this study was constructed from fracture-surface images mined from open-access literature under multiple licenses, including CC0, CC BY, CC BY-ND, CC BY-NC, CC BY-SA, CC BY-NC-SA, and CC BY-NC-ND. To respect license restrictions, particularly NonCommercial and NoDerivatives conditions, we do not publicly redistribute the processed image dataset, image crops, contrast-normalized images, cached training shards, or other pixel-level derivatives from restricted sources. Instead, we provide source metadata, license annotations, and preprocessing scripts to support scientific reproducibility. For materials licensed under CC BY-ND 4.0 and CC BY-NC-ND 4.0, preprocessing operations such as cropping, panel extraction, filtering, resizing, and contrast normalization were performed solely for internal academic research, including training and evaluation; for CC BY-NC-ND 4.0 materials, this use was further restricted to noncommercial purposes. No original or adapted CC BY-ND or CC BY-NC-ND image files are included in the public release.

The released model files are fine-tuned weights for Qwen3-VL-32B-Instruct, an image-text-to-text and text-to-text vision-language model used here for visual understanding and textual prediction tasks, not pixel-level image generation. The released model files contain no image files, thumbnails, cached training examples, cached embeddings, dataloaders, or other redistributed pixel-level derivatives of the source literature. Because the released files are model parameter tensors rather than image data, we expect the risk of reconstructing source images from the weights to be low [63]. We further evaluated this risk through memorization and leakage checks, including prompt-based attempts to elicit memorized captions or figure-specific descriptions and similarity checks against source images. Under our evaluation protocol, we did not observe reproduction or reconstruction of source figures.

The fine-tuned model weights are released for noncommercial research use only. We view this work as noncommercial academic research involving text-and-data-mining-style use of open-access scientific literature. Because copyright exceptions, fair use, and the application of Creative Commons license conditions to AI training are jurisdiction- and fact-specific, the public release should not be interpreted as redistributing CC BY-ND or CC BY-NC-ND image material or as authorizing commercial use.

**Table 1. Comprehensive benchmarking results for the Qwen3-VL-32B-Instruct, FT-Qwen, GPT-5.5-Reasoning, GPT-5.5-Reasoning + FT-Qwen, Gemini 3.1 Pro-Reasoning, and Gemini 3.1 Pro-Reasoning + FT-Qwen.** O denotes the **Original** training set, R denotes the **Rebalanced** training set, N denotes the **Non-canonical** output format, and C denotes the **Canonical** output format. Pre denotes precision, Rec denotes recall.

| Model | Training set | | Output format | Hold-out test set | | | Manual test set | | |
|---|---|---|---|---|---|---|---|---|---|
| | | | | Pre | Rec | F1 | Pre | Rec | F1 |
| Qwen3-VL-32B-Instruct | – | | N | 0.36 | 0.77 | 0.49 | 0.35 | 0.68 | 0.46 |
| | | | C | 0.35 | 0.36 | 0.35 | 0.29 | 0.30 | 0.29 |
| FT-Qwen | w/ aug & w/ extra collection | O | N | 0.81 | 0.74 | 0.77 | **0.92** | 0.88 | 0.90 |
| | | | C | 0.81 | 0.76 | 0.78 | 0.85 | 0.82 | 0.83 |
| | | R | N | 0.77 | 0.75 | 0.76 | 0.88 | 0.85 | 0.86 |
| | | | C | **0.83** | 0.80 | **0.81** | 0.86 | 0.83 | 0.84 |
| | w/ aug & w/o extra collection | O | N | 0.72 | 0.69 | 0.70 | 0.81 | 0.77 | 0.79 |
| | | | C | 0.71 | 0.68 | 0.69 | 0.77 | 0.71 | 0.74 |
| | | R | N | 0.76 | 0.73 | 0.74 | 0.80 | 0.77 | 0.78 |
| | | | C | 0.74 | 0.73 | 0.73 | 0.82 | 0.78 | 0.80 |
| | w/o aug & w/ extra collection | O | N | 0.76 | 0.71 | 0.73 | 0.83 | 0.79 | 0.81 |
| | | | C | 0.73 | 0.69 | 0.71 | 0.84 | 0.83 | 0.83 |
| | | R | N | 0.75 | 0.74 | 0.74 | 0.82 | 0.80 | 0.81 |
| | | | C | 0.76 | 0.75 | 0.75 | 0.80 | 0.78 | 0.79 |
| | w/o aug & w/o extra collection | O | N | 0.71 | 0.67 | 0.69 | 0.72 | 0.67 | 0.69 |
| | | | C | 0.67 | 0.64 | 0.65 | 0.75 | 0.70 | 0.72 |
| | | R | N | 0.71 | 0.71 | 0.71 | 0.71 | 0.70 | 0.70 |
| | | | C | 0.75 | 0.74 | 0.74 | 0.72 | 0.71 | 0.71 |
| GPT-5.5-Reasoning | – | | N | 0.54 | 0.77 | 0.63 | 0.58 | 0.80 | 0.67 |
| | | | C | 0.56 | 0.71 | 0.63 | 0.62 | 0.75 | 0.68 |
| | – | | N | 0.77 | 0.80 | 0.78 | 0.85 | 0.90 | 0.87 |

| GPT-5.5-Reasoning + FT-Qwen | | C | 0.80 | 0.80 | 0.80 | 0.83 | 0.86 | 0.84 |
|---|---|---|---|---|---|---|---|---|
| Gemini 3.1 Pro-Reasoning | – | N | 0.64 | 0.78 | 0.70 | 0.78 | 0.85 | 0.81 |
| | | C | 0.63 | 0.73 | 0.68 | 0.77 | 0.85 | 0.81 |
| Gemini 3.1 Pro-Reasoning + FT-Qwen | – | N | 0.80 | **0.82** | **0.81** | **0.92** | **0.92** | **0.92** |
| | | C | 0.79 | 0.81 | 0.80 | 0.88 | 0.88 | 0.88 |

Supplementary Materials for

# Fine-tuning a vision-language model for fracture-surface morphology recognition

Quanliang Liu *et al.*

Corresponding author: Hyunseok Oh (hyunseok.oh@wisc.edu)

**This PDF file includes:**

# Contents

**Note S1. Definitions of the 11 fracture-surface features.**

All following images are adapted with CC BY permission.

### S1.1 Dimple, inclusion particle [1]

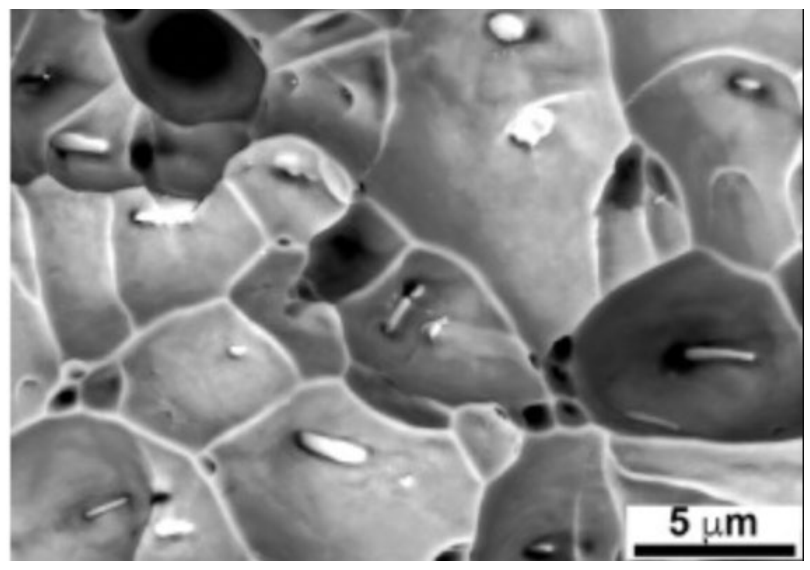

Dimple: cup-like pits from microvoid coalescence.

Inclusion particle: particles clearly embedded within the matrix or fracture surface. Do not use this classification for loose debris, particles merely resting on the surface, or ambiguous features.

### S1.2 Striation [2]

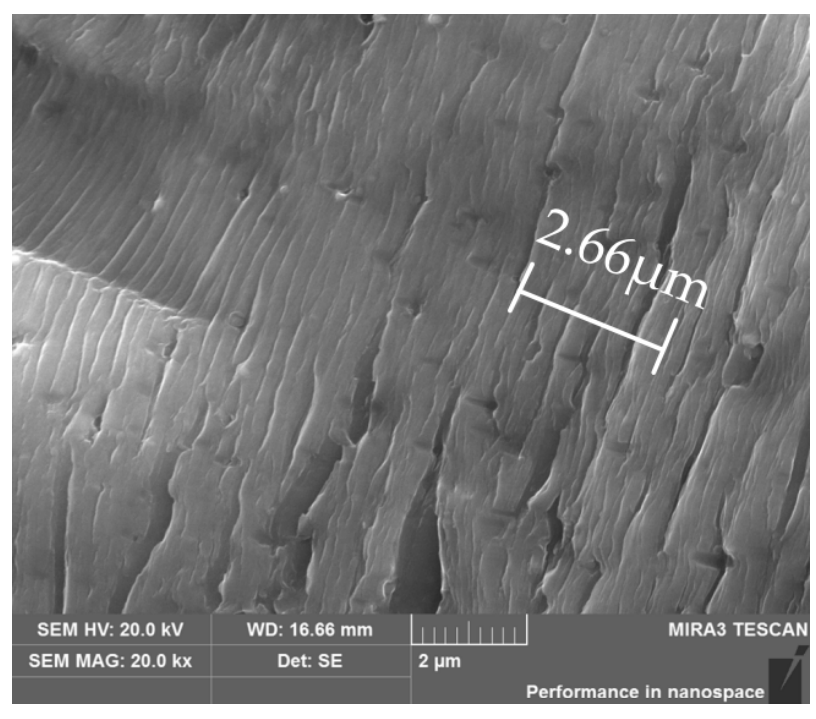

Striation: fine parallel fatigue lines.

### S1.3 Intergranular facet [3]

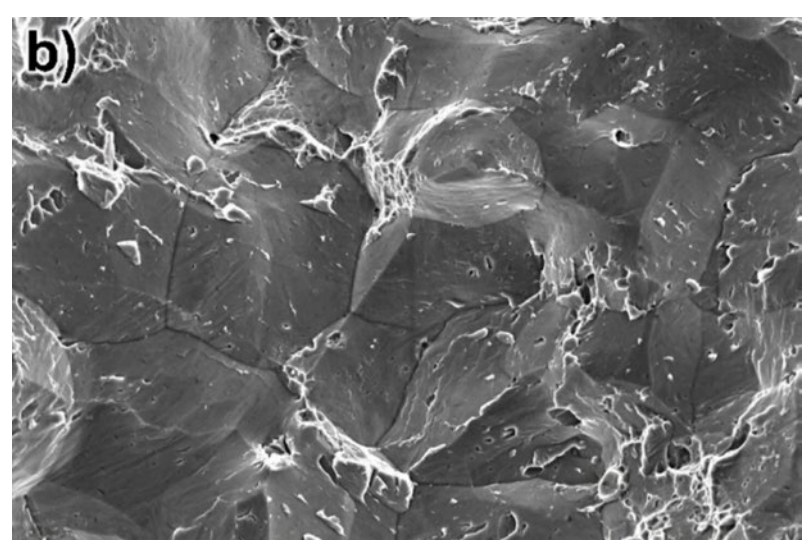

Intergranular facet: faceted appearance following grain boundaries.

### S1.4 Shear lip [4]

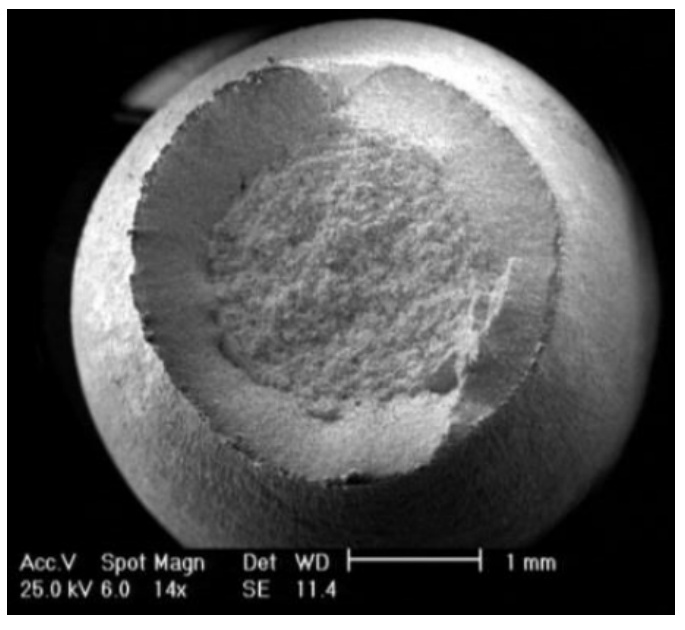

Shear lip: slanted edge zone from final shear deformation.

### S1.5 Ratchet mark [5]

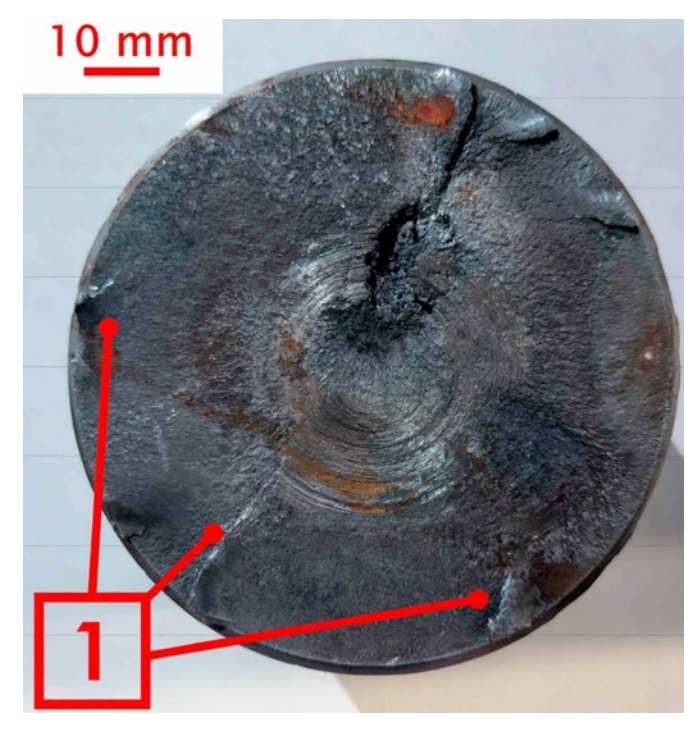

Ratchet mark: step-like ridges from coalescing cracks.

### S1.6 Cleavage, river pattern [6]

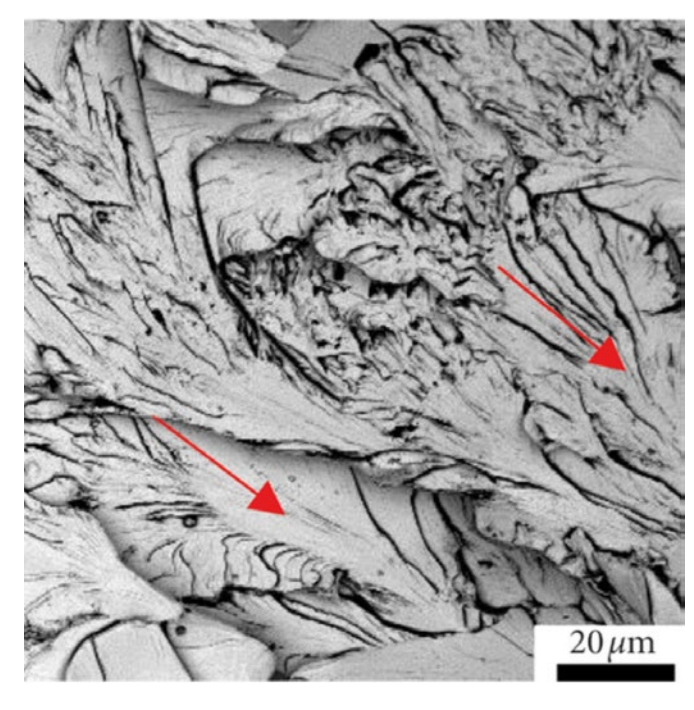

Cleavage: relatively flat facets with step-like changes in height.

River pattern: converging step-like lines on cleavage facets.

### S1.7 Fish-eye [7]

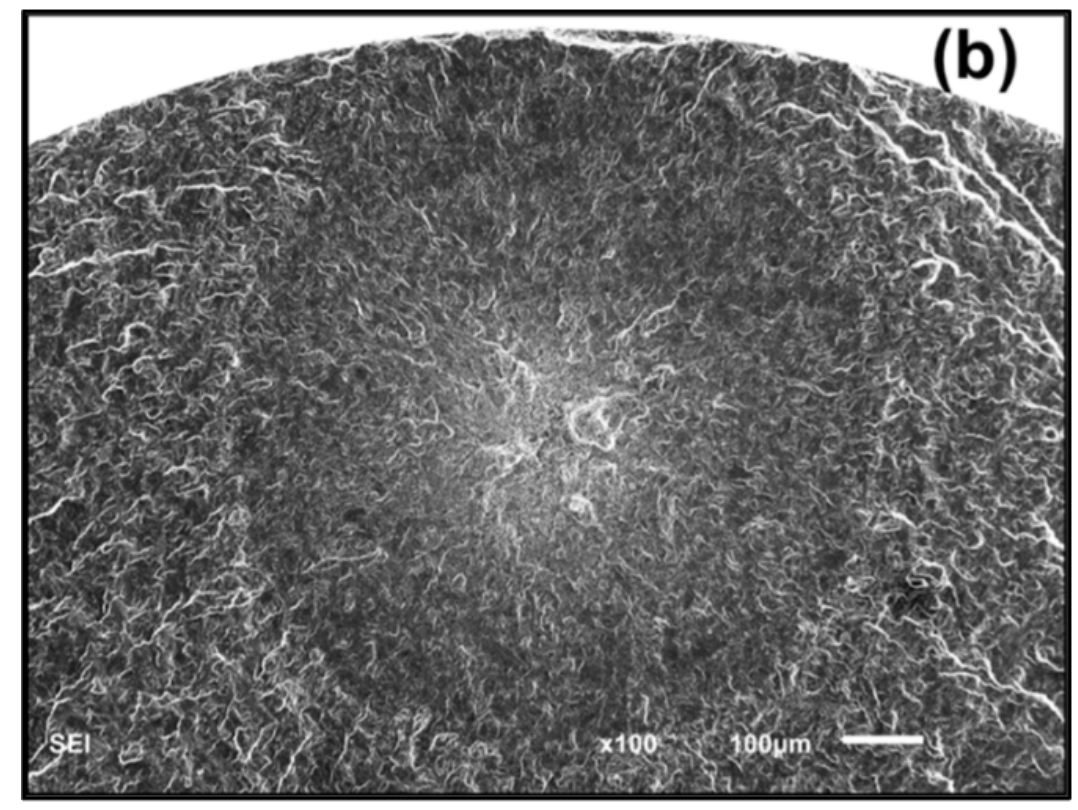

Fish-eye: circular or elliptical interior fatigue region.

### S1.8 Chevron [8]

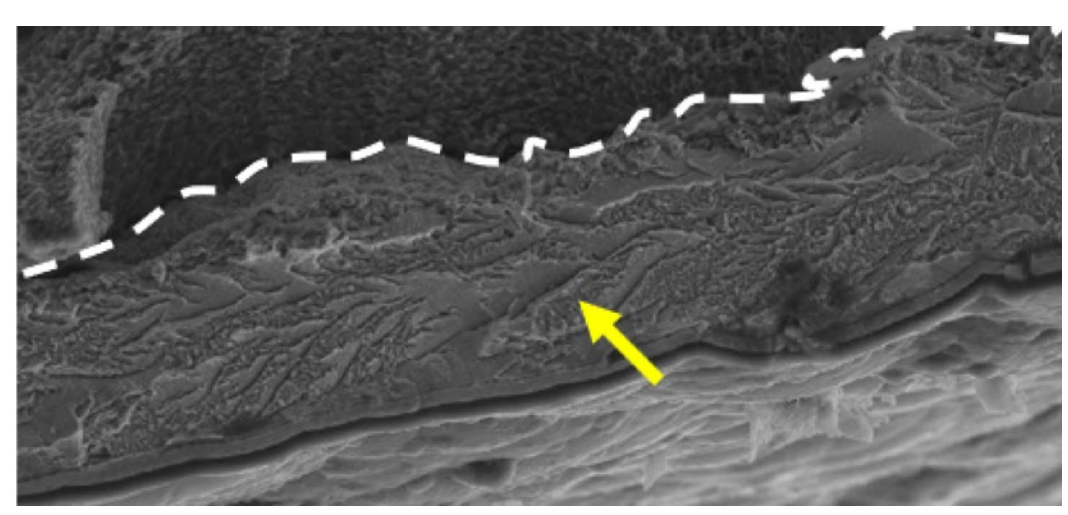

Chevron: V-shaped marks pointing toward the origin.

### S1.9 Beach mark [9]

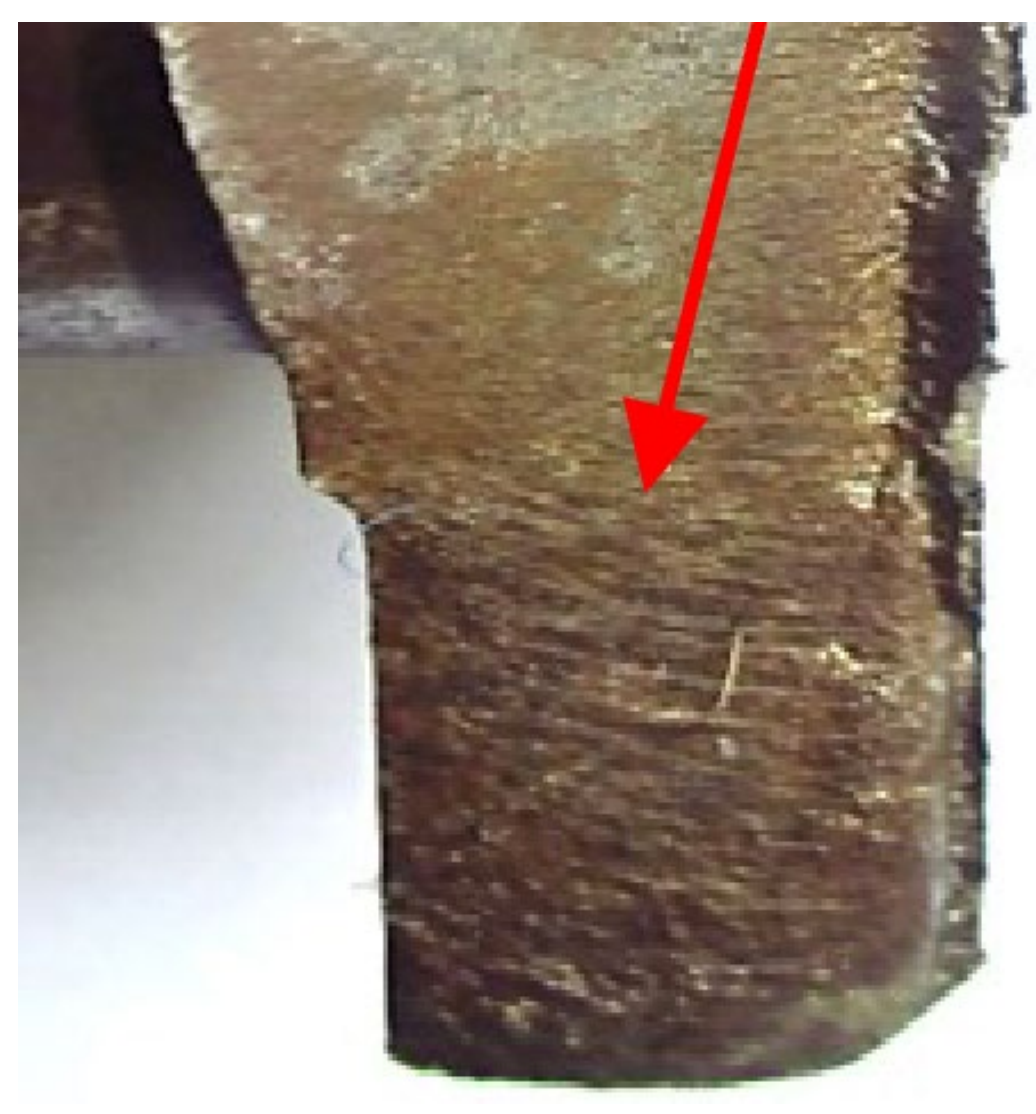

Beach mark: curved or concentric macroscopic arrest / progression lines.

## Note S2. Prompts used for annotation, fine-tuning, inference, specialist-assisted proprietary models analysis, proprietary–specialist integration in fatigue fracture analysis.

This note collects the prompts for dataset annotation and quality control, fine-tuning and inference of models, specialist-assisted inference for the proprietary models, and the proprietary–specialist integration in fatigue fracture analysis.

### S2.1 Annotation prompt

For each single-panel image, GPT-5.2-Reasoning (high) [10] was given the image, the identifier of the image (with panel number), a relevant excerpt from the source paper, and a structured two-part prompt: a system prompt and a user prompt. For manually collected images, features were assigned by human annotators and GPT-5.2-Reasoning was used only to produce the morphology description.

**System prompt:**

> You are an expert fractographer annotating fracture-surface image panels for a dataset used in VLM fine-tuning.
>
> Authority + eligibility (panel-locked, two-stage):
>
> 1) TEXT ALLOW-LIST: Only paper-excerpt sentences that explicitly refer to this exact panel make a feature class eligible to mention.
>
> 2) LOW-INFO FALLBACK (only if needed): If those eligible sentences mention no feature classes (only generic fracture-mode terms like brittle/ductile/fatigue/overload, or no morphology), you may infer feature classes from the panel itself—but only classes that are unmistakably visible at this panel's scale/resolution. If unsure, omit.
>
> Visibility gate (always):
>
> - Even if a feature class is stated in eligible text, you may mention/tag it only if it is clearly observable/identifiable in this panel at its scale/resolution. If not confidently verifiable here, omit it.
>
> Empty result:
>
> - If no feature classes are both eligible (by text, or fallback when allowed) AND visibly verifiable in this panel, output exactly: N/A
>
> General constraints:
>
> - Do not assume material, alloy, environment, or loading history unless explicitly stated in eligible excerpt sentences.
>
> - Do not infer failure mechanism beyond the short cue definitions below.
>
> - Use standard fractography terminology.
>
> Feature classes (use these exact names only):

dimple, cleavage, striation, shear lip, beach mark, fish-eye, ratchet mark, intergranular facet, river pattern, chevron, inclusion particle

Definitions (short cues for interpretation of stated features only):

- dimple: cup-like pits from microvoid coalescence.
- cleavage: relatively flat facets with step-like changes in height.
- striation: fine parallel fatigue lines.
- shear lip: slanted edge zone from final shear deformation.
- beach mark: curved or concentric macroscopic arrest/progression lines.
- fish-eye: circular or elliptical interior fatigue region.
- ratchet mark: step-like ridges from coalescing cracks.
- intergranular facet: faceted appearance following grain boundaries.
- river pattern: converging step-like lines on cleavage facets.
- chevron: V-shaped marks pointing toward the origin.
- inclusion particle: particles clearly embedded within the matrix or fracture surface. Do not use this classification for loose debris, particles merely resting on the surface, or ambiguous features.

Scale labels:

- Macroscopic: whole fracture regions / large markings (e.g., beach mark, chevron, ratchet mark).
- Microscopic: micro-topography (e.g., striation, river pattern).

**User prompt:**

Relevant paper excerpt:{paper_excerpt}

Image metadata: {image_panel_identifier}

Panel-lock:

Use ONLY excerpt sentences that explicitly refer to this panel. Ignore text about other figures/panels/samples/conditions.

Decision process:

1) From eligible excerpt sentences, list the feature classes explicitly stated as present (or unambiguous equivalents). These are candidates.

2) Visibility check: keep only candidate classes that are clearly observable/identifiable in this panel at its scale/resolution.

3) Low-info fallback (only if step 1 yields no feature classes): infer from the panel only those classes from the allowed list that are unmistakably visible. If unsure, omit.

4) If no classes remain after steps 2–3, output exactly: N/A

5) Otherwise output exactly two parts:

- One paragraph describing ONLY the remaining classes (and only what is visible here). You may localize where they appear and choose Macroscopic vs Microscopic.

- On a new line: the remaining feature classes, comma-separated, using the exact class names.

Writing constraints:

- Begin the paragraph with "Macroscopic:" or "Microscopic:" (mm-scale or larger = Macroscopic; otherwise Microscopic).

- Do not add any feature, texture, mechanism, or detail not both (eligible) and (visibly verifiable) under the rules.

- Avoid abbreviations/acronyms.

- Do not use meta references like "the paper," "the excerpt," "the authors," or "the image."

- Neutral, observational tone.

Output format (nothing else):

Option A: N/A

Option B: 1) Single paragraph. 2) Feature classes line (comma-separated).

### S2.2 Quality-control filtering prompt

GPT-5.4-Reasoning (high) [11] was used as a second-stage filter to remove images whose fracture-surface appearance did not clearly support the annotation produced in **Section S2.1**. The filter kept only panels for which GPT-5.4-Reasoning returned "clear"; panels returning "unclear" were dropped.

**User prompt:**

You are analyzing a fracture surface image.

Task: Decide whether the visible fracture-surface appearance clearly supports the current morphology annotation.

Current morphology annotation:{morphology}

Rules:

- Evaluate only whether the CURRENT morphology annotation is visually supported.

- Ignore arrows, boxes, labels, and any overlaid text.

- Base your judgment only on visible fracture-surface evidence in the image.

- If the image is blurry, low-contrast, ambiguous, or does not clearly show evidence for the current morphology annotation, return "unclear".

- Return "clear" only when the visible fracture surface supports the current morphology annotation.

- Do NOT rewrite, correct, remove, or add morphology labels.

Output format:

Part 1: Return exactly one word: clear or unclear

Part 2: Briefly explain whether the visible fracture-surface appearance supports the current morphology annotation.

### S2.3 Fine-tuning and inference prompts

Two output schemas were used during fine-tuning and inference: a non-canonical schema that allows a short morphological rationale followed by a free-form list of predicted features, and a canonical schema that forces the model to emit a fixed 11-key JSON object.

**Non-canonical prompt:**

You are given a fracture surface image. Let's think step by step. First, determine if the image is macroscopic or microscopic. Next, describe the visible fracture morphology in detail. Finally, identify the morphological features present from the following list (select all that apply): dimple, cleavage, river pattern, intergranular facet, shear lip, striation, beach mark, ratchet mark, fish-eye, chevron, inclusion particle.

Completely ignore arrows, boxes, and text drawn on the image; rely only on visual evidence.

Output format:

<think>Scale (Macroscopic/Microscopic): morphology description</think>

<answer>feature 1, feature 2, ...</answer>

**Canonical prompt:**

You are given a fracture surface image. Let's think step by step. First, determine if the image is macroscopic or microscopic. Next, describe the visible fracture morphology in detail. Finally, set the following feature flags to 1 if the feature is present, otherwise 0: dimple, cleavage, river pattern, intergranular facet, shear lip, striation, beach mark, ratchet mark, fish-eye, chevron, inclusion particle.

Completely ignore arrows, boxes, and text drawn on the image; rely only on visual evidence.

Output format:

<think>Scale (Macroscopic/Microscopic): morphology description</think>

<answer "dimple":0, "cleavage":0, "river pattern":0, "intergranular facet":0, "shear lip":0, "striation":0, "beach mark":0, "ratchet mark":0, "fish-eye":0, "chevron":0, "inclusion particle":0}</answer>

(Use only these keys exactly. Output valid JSON in the <answer> block.)

### S2.4 Specialist-assisted inference for proprietary models

In the assisted setting, the output from the fine-tuned specialist model was appended to the base prompt as the following suffix before the query was sent to GPT-5.5-Reasoning [12] or Gemini 3.1 Pro-Reasoning [13].

Additional reference from a specialized fine-tuned model:

{HINT FROM SPECIALIST}

This is NOT the ground truth. Use it as a secondary signal only.

### S2.5 Proprietary–specialist integration in fatigue fracture analysis

In this case, the specialist was first run on each of the four images (Image 1, Image 2, Image 2-1, Image 2-2) using the non-canonical prompt. The four specialist responses were then packaged with the images and passed to GPT-5.5-Reasoning under the comparative prompt below, followed by the specialist-assistance suffix from **S2.4**.

I have provided four SEM images of fracture surfaces from the same FCC HEA tested under two different conditions. Image 1 corresponds to the first condition. Image 2, Image 2-1, and Image 2-2 correspond to the second condition. Image 2 is a broader view of the second condition and contains an arrow marking a specific region of interest. Image 2-1 is a zoomed-in view of that marked region, and Image 2-2 is a further zoomed-in view of the same region shown in Image 2-1. Treat Image 2, Image 2-1, and Image 2-2 as complementary views of the SAME second condition, not as separate samples. Use the arrow only to understand which region is being magnified across images, not as direct evidence of fracture morphology. Relying strictly on the fracture morphology and the provided scale bars, hypothesize the environmental or testing conditions responsible for the fracture modes seen in Image 1 versus the second-condition views (Image 2, Image 2-1, and Image 2-2).

Additionally, detail the factors that likely caused the fracture mechanism to shift from the first condition to the second condition.

## Note S3. Fine-tuning training settings

Fine-tuning was performed using the LLaMA-Factory [14] framework with low-rank adaptation (LoRA [15]) on four NVIDIA H100 GPUs on the Purdue Anvil cluster [16]. The base model was Qwen3-VL-32B-Instruct [17]. Identical settings were used for every run; the 16 training configurations differ only along four binary axes—image augmentation (on / off), supplementary image collection (included / not included), output schema (canonical vs non-canonical), and feature-balance downsample (original vs rebalanced)—yielding 2 × 2 × 2 × 2 cases. The following are the detailed training configurations.

**Backbone:** Qwen/Qwen3-VL-32B-Instruct

**Template:** qwen3_vl_nothink

**Training stage:** SFT

**Adaptation method:** LoRA

**LoRA target modules:** all linear layers

**LoRA rank / alpha / dropout:** 64 / 128 / 0.1

**Precision:** bfloat16

**Gradient checkpointing:** enabled

**Per-device train / eval batch size:** 1 / 1

**Gradient accumulation steps:** 8

**Learning rate / schedule:** 1e-5, cosine

**Warmup ratio:** 0.03

**Epochs:** 4

**Max sequence length (cutoff_len):** 4096 tokens

**Validation split (val_size):** 5%

**Evaluation strategy:** steps (eval_steps = 50)

**Checkpoint strategy:** steps (save_steps = 200, keep last 2)

**Logging:** every 10 steps, TensorBoard

**Inference decoding:** greedy

### Table S1. Statistics of the hold-out and manual test sets

Two test sets were used to benchmark all models. The hold-out test set comprises 100 images selected from the constructed dataset with a balanced feature distribution. The manual test set comprises 100 additional images that are absent from the constructed dataset; these images were manually collected and labeled, with the feature distribution also balanced. Table values represent the image count for each fracture feature across both test sets.

| Feature class | Hold-out test set | Manual test set |
|---|---|---|
| dimple | 11 | 13 |
| cleavage | 15 | 14 |
| striation | 11 | 11 |
| beach mark | 12 | 12 |
| ratchet mark | 11 | 13 |
| fish-eye | 11 | 11 |
| inclusion particle | 11 | 9 |
| intergranular facet | 11 | 11 |
| river pattern | 11 | 13 |
| shear lip | 11 | 11 |
| chevron | 11 | 10 |

**Table S2. Statistics of the three ablation training sets.**

The main text reports a fully enriched training configuration (with rotation augmentation and with targeted extra collection). The three ablation cases remove one or both ingredients: augmented-only (with augmentation, without extra collection), extra-collection-only (without augmentation, with extra collection), and minimal (without augmentation, without extra collection). For each case, both the original and rebalanced variants are reported; the rebalanced split downsamples dimple and cleavage. O denotes the original training set, R denotes the rebalanced training set.

| Feature class | w/ aug & w/o extra | | w/o aug & w/ extra | | w/o aug & w/o extra | |
|---|---|---|---|---|---|---|
| | O | R | O | R | O | R |
| dimple | 7,162 | 567 | 7,150 | 472 | 7,147 | 472 |
| cleavage | 2,357 | 567 | 2,372 | 472 | 2,339 | 472 |
| striation | 567 | 567 | 223 | 223 | 138 | 138 |
| beach mark | 245 | 245 | 188 | 188 | 56 | 56 |
| ratchet mark | 223 | 223 | 140 | 140 | 52 | 52 |
| fish-eye | 90 | 90 | 124 | 124 | 21 | 21 |
| inclusion particle | 299 | 299 | 341 | 341 | 269 | 269 |
| intergranular facet | 475 | 475 | 472 | 472 | 472 | 472 |
| river pattern | 384 | 384 | 411 | 411 | 384 | 384 |
| shear lip | 365 | 365 | 356 | 356 | 350 | 350 |
| chevron | 130 | 130 | 64 | 64 | 28 | 28 |
| **Total images** | **11,217** | **3,106** | **10,761** | **2,467** | **10,269** | **2,006** |

### Table S3. Decoding-sweep configurations for proprietary–specialist integration in fatigue fracture analysis.

To verify that the specialist-assisted identifications in **Fig. 8** were stable under non-greedy decoding, the specialist model was run on all four panels (Image 1, Image 2, Image 2-1, Image 2-2) across ten decoding configurations: one greedy baseline and nine temperature/top-p combinations. The resulting specialist responses were then passed, together with the four images, to GPT-5.5-Reasoning.

| Run ID | Mode | Temperature | Top-p | Purpose |
|---|---|---|---|---|
| run_00_greedy | greedy | – | – | baseline |
| run_01_t0.3_p0.9 | sampling | 0.3 | 0.9 | low temperature, standard top-p |
| run_02_t0.5_p0.9 | sampling | 0.5 | 0.9 | moderate temperature |
| run_03_t0.7_p0.9 | sampling | 0.7 | 0.9 | reference temperature |
| run_04_t0.9_p0.9 | sampling | 0.9 | 0.9 | high temperature |
| run_05_t0.7_p0.3 | sampling | 0.7 | 0.3 | narrow top-p |
| run_06_t0.7_p0.7 | sampling | 0.7 | 0.7 | moderate top-p |
| run_07_t0.7_p0.95 | sampling | 0.7 | 0.95 | wide top-p |
| run_08_t0.7_p0.9_rep2 | sampling | 0.7 | 0.9 | repeat of run_03 |
| run_09_t0.7_p0.9_rep3 | sampling | 0.7 | 0.9 | repeat of run_03 |

## Figure S1. Loss curves — with augmentation, with extra collection.

Training loss (left) and evaluation loss (right) across training steps for all four cases.

**Canonical output schema, original training set**

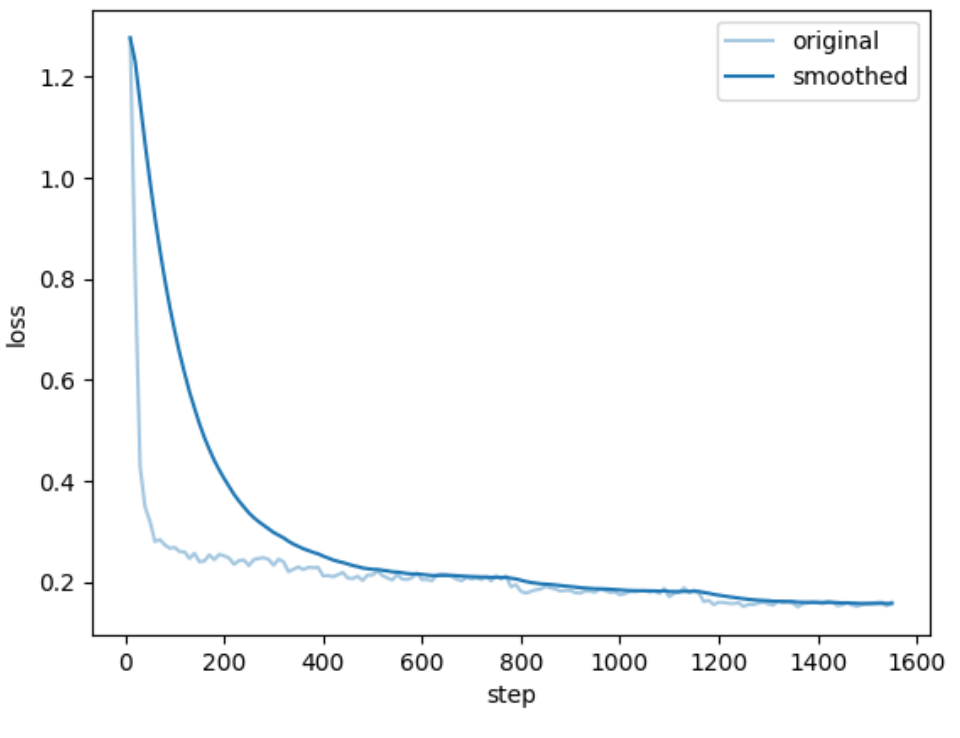

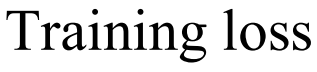

Training loss

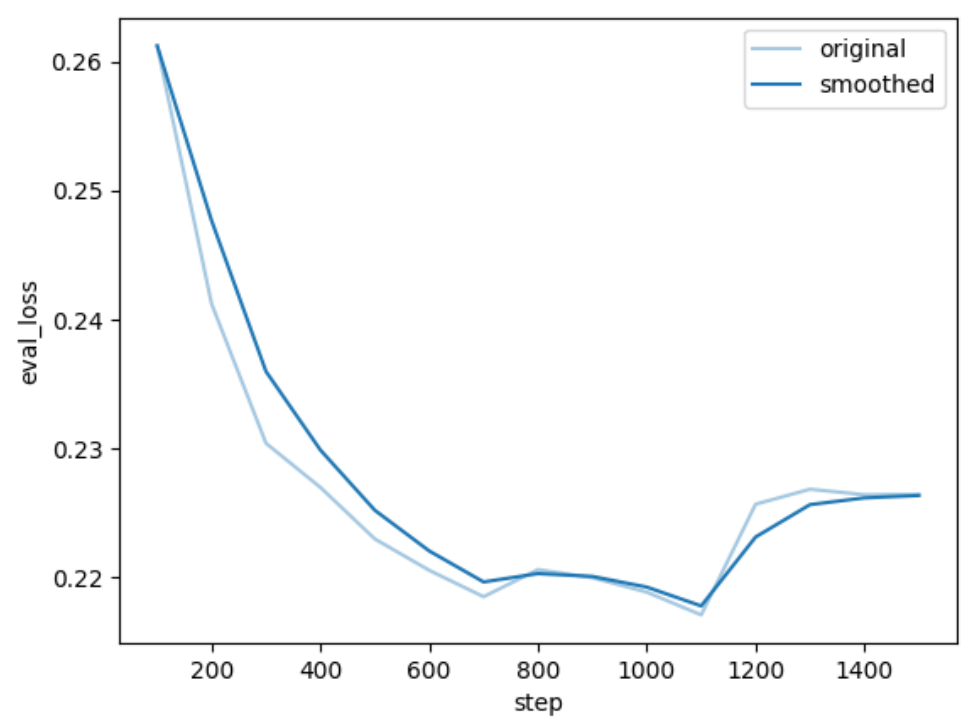

Evaluation loss

**Canonical output schema, rebalanced training set**

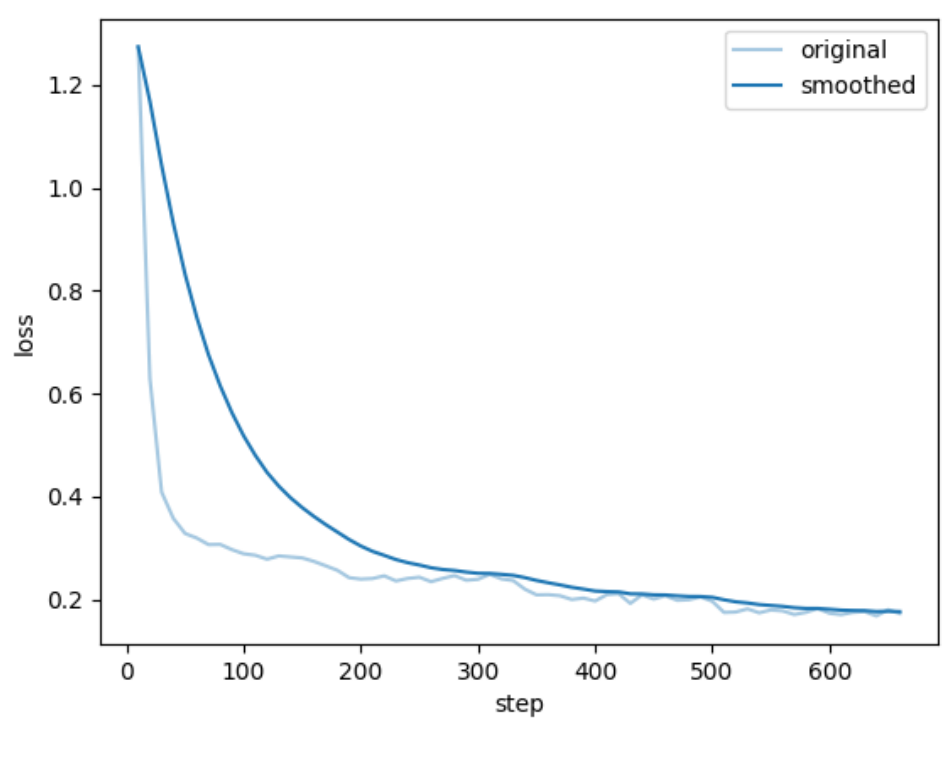

Training loss

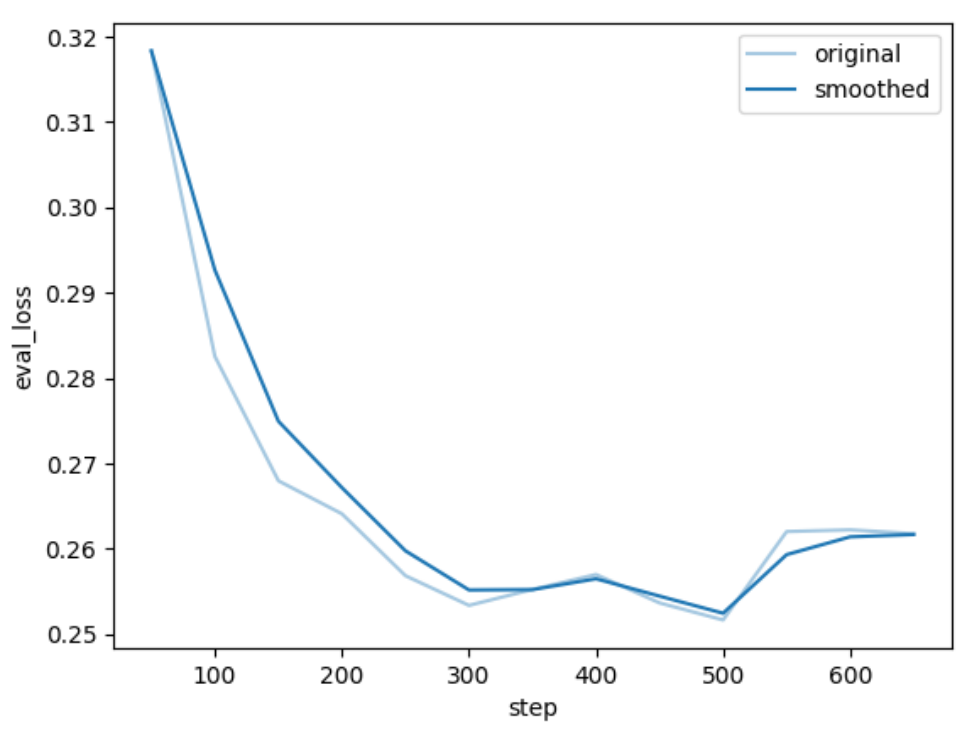

Evaluation loss

**Non-canonical output schema, original training set**

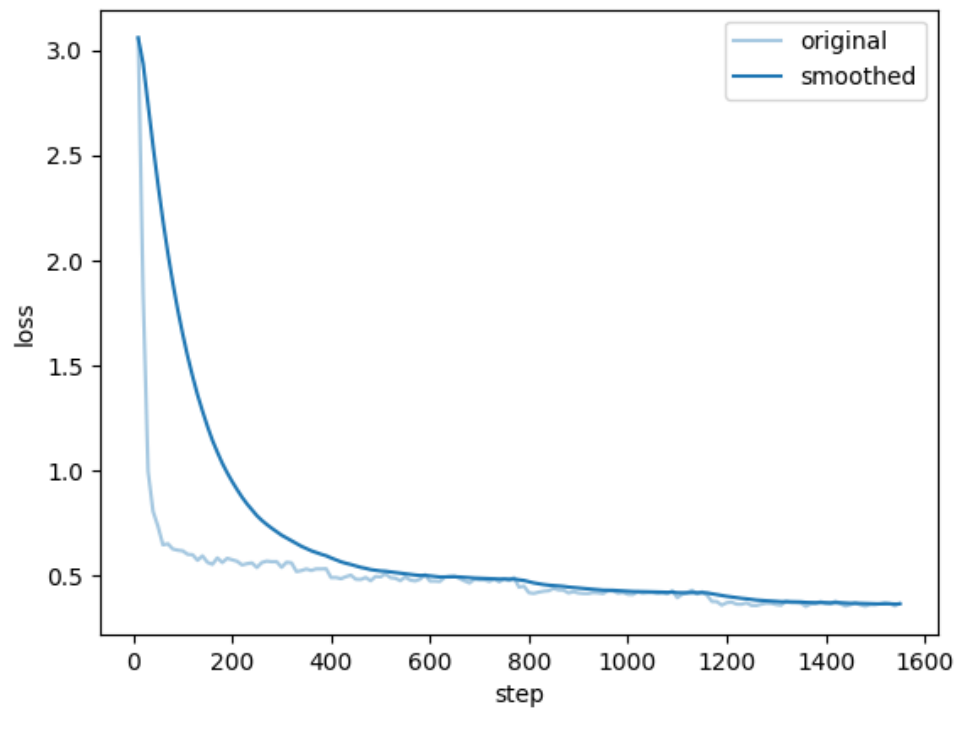

Training loss

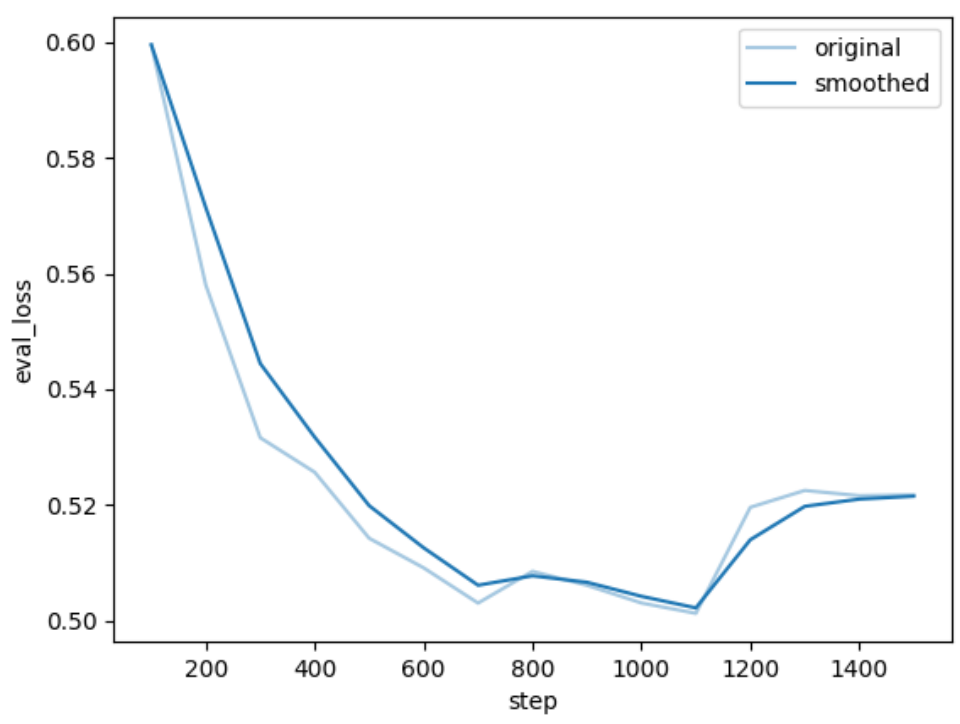

Evaluation loss

**Non-canonical output schema, rebalanced training set**

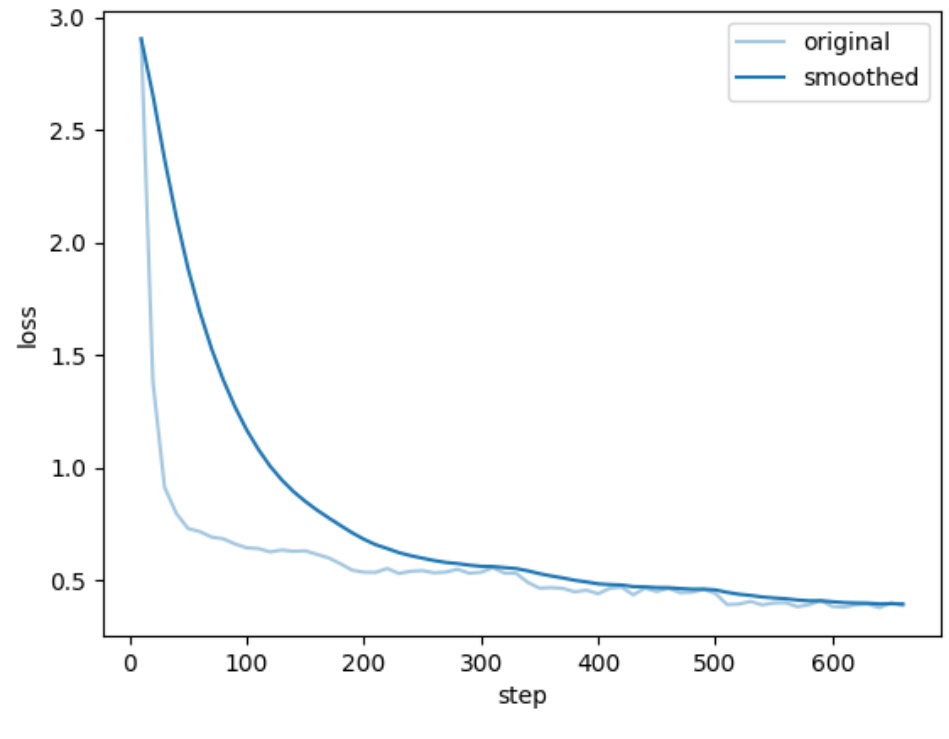

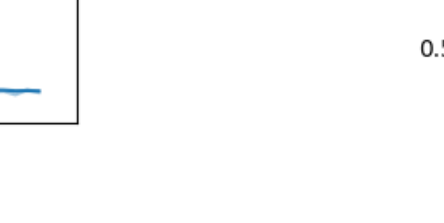

Training loss

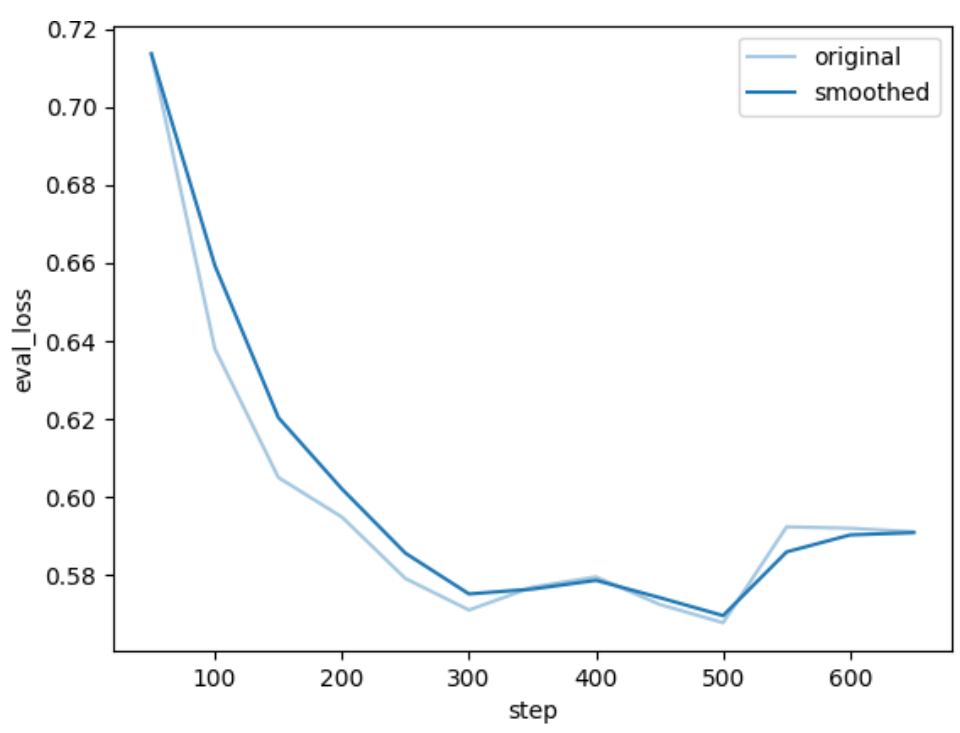

Evaluation loss

## Figure S2. Loss curves — with augmentation, without extra collection.

Training loss (left) and evaluation loss (right) across training steps for all four cases.

### Canonical output schema, original training set

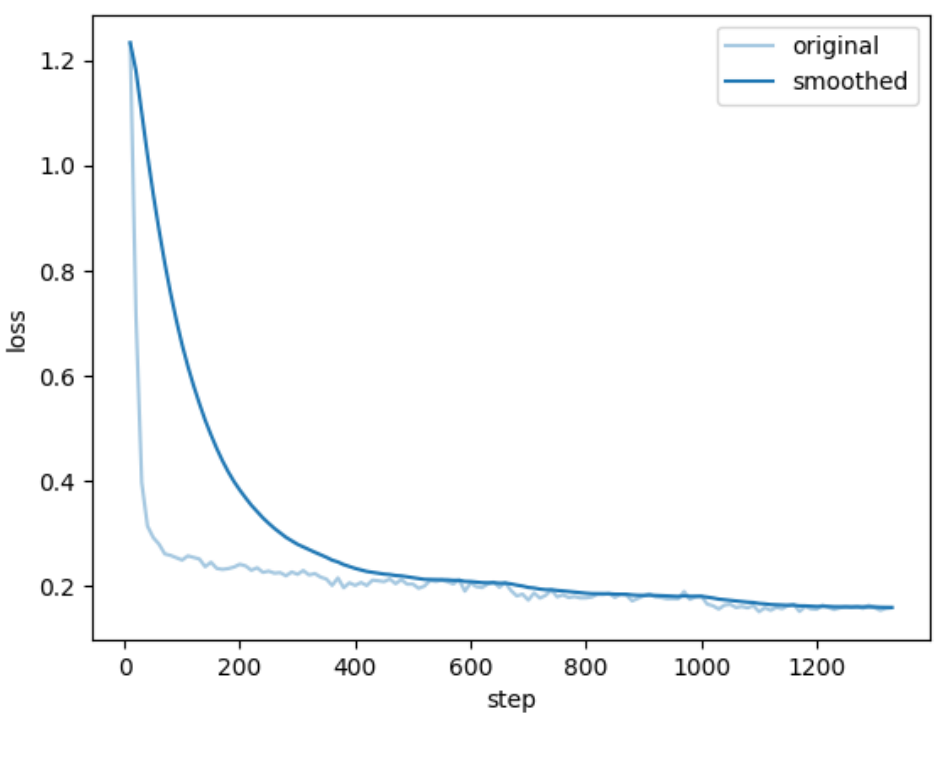

Training loss

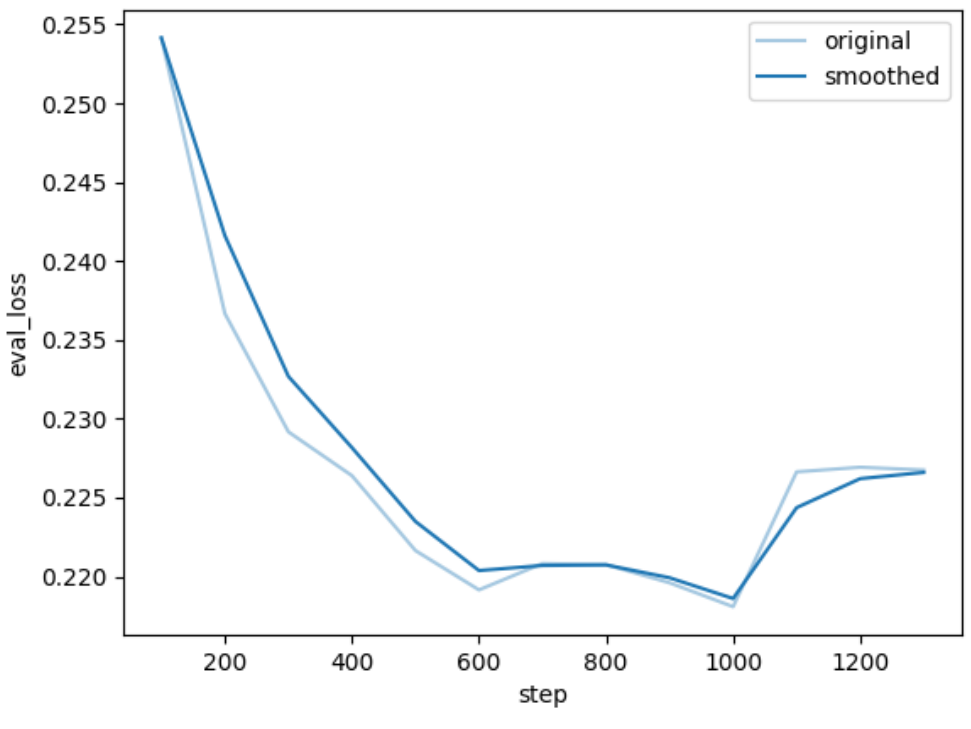

Evaluation loss

### Canonical output schema, rebalanced training set

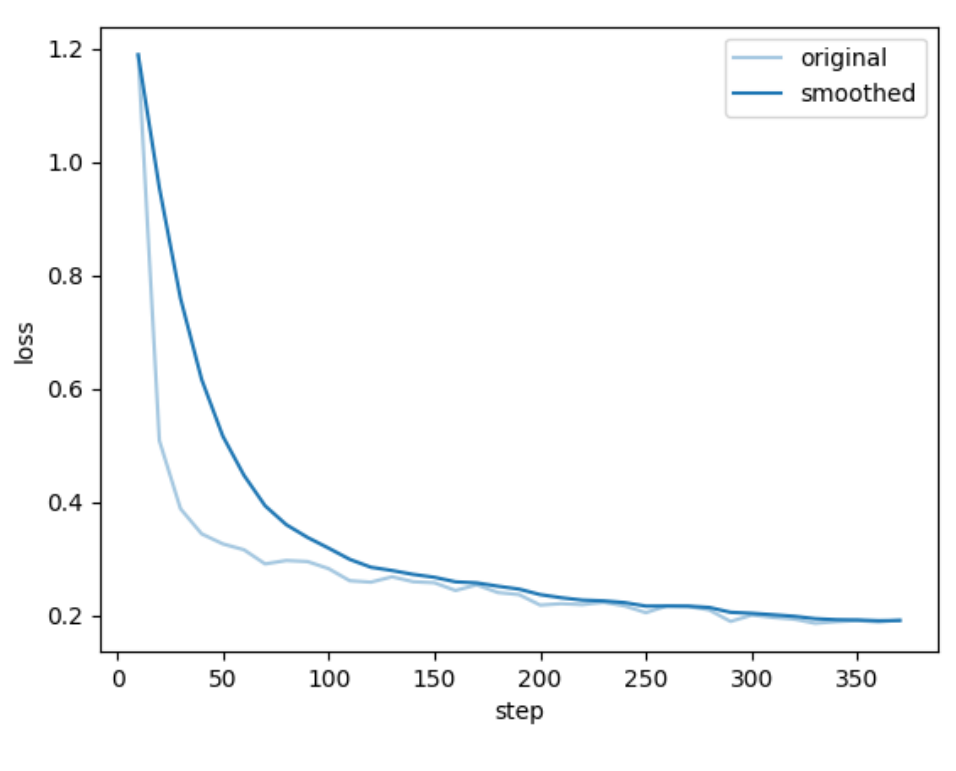

Training loss

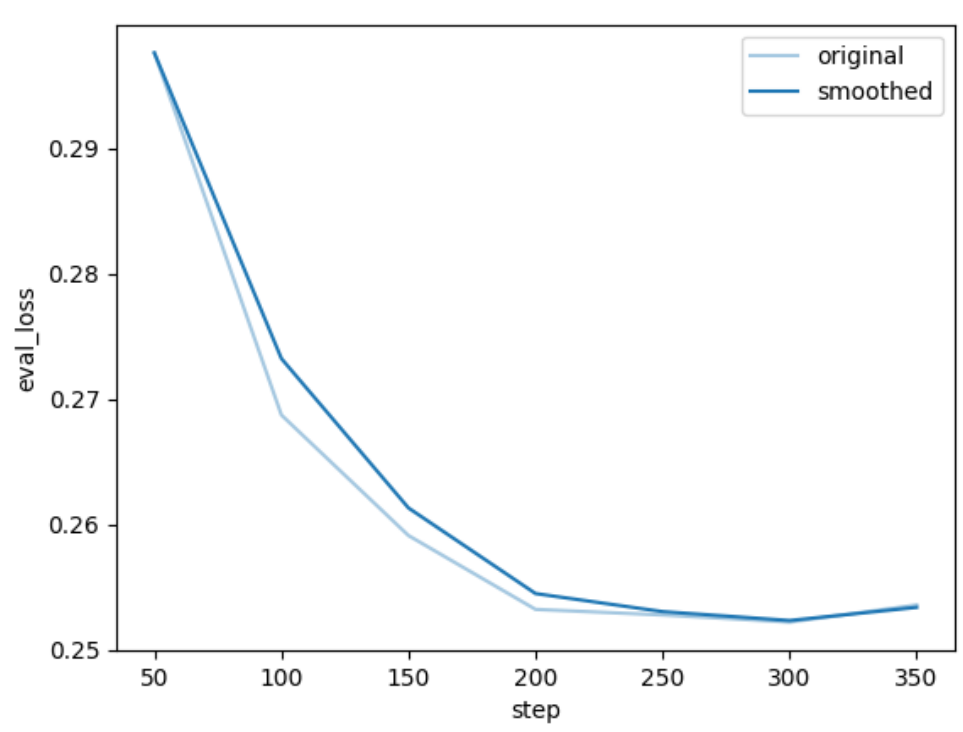

Evaluation loss

### Non-canonical output schema, original training set

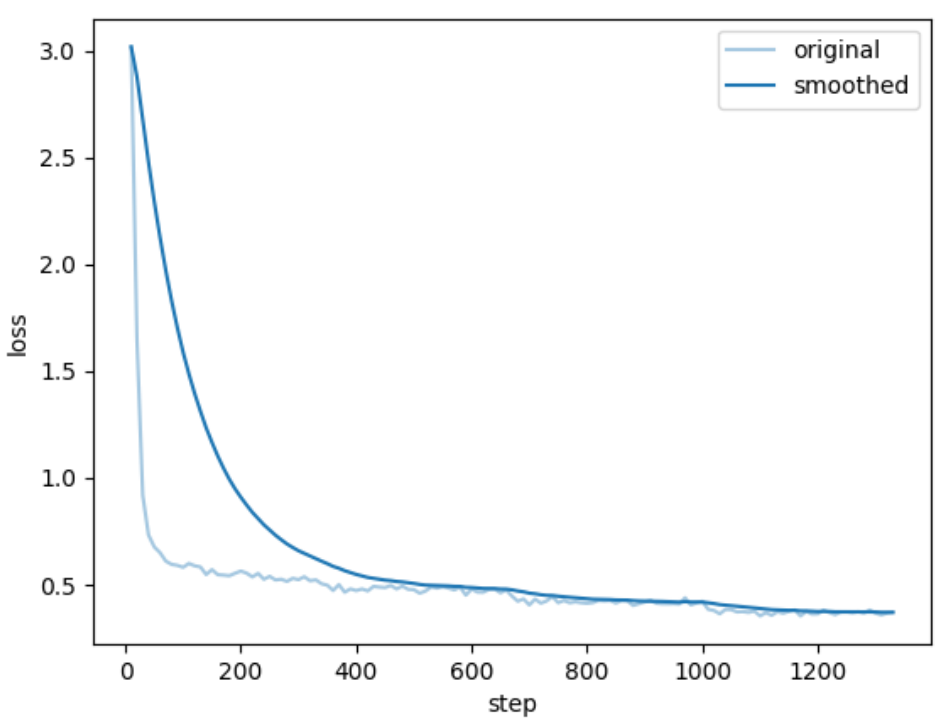

Training loss

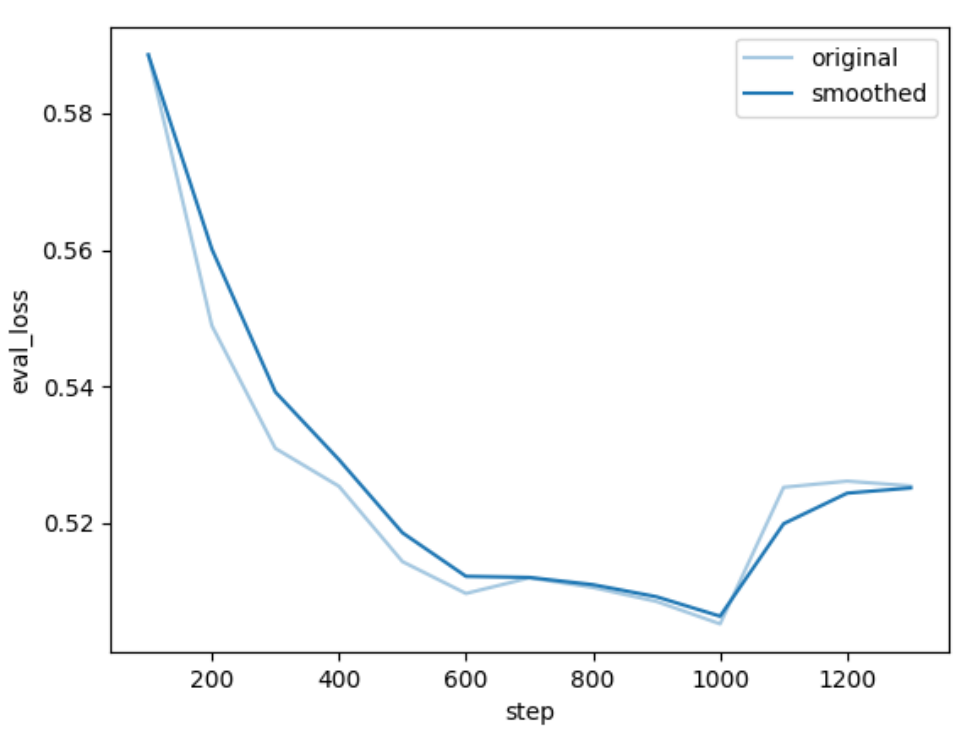

Evaluation loss

**Non-canonical output schema, rebalanced training set**

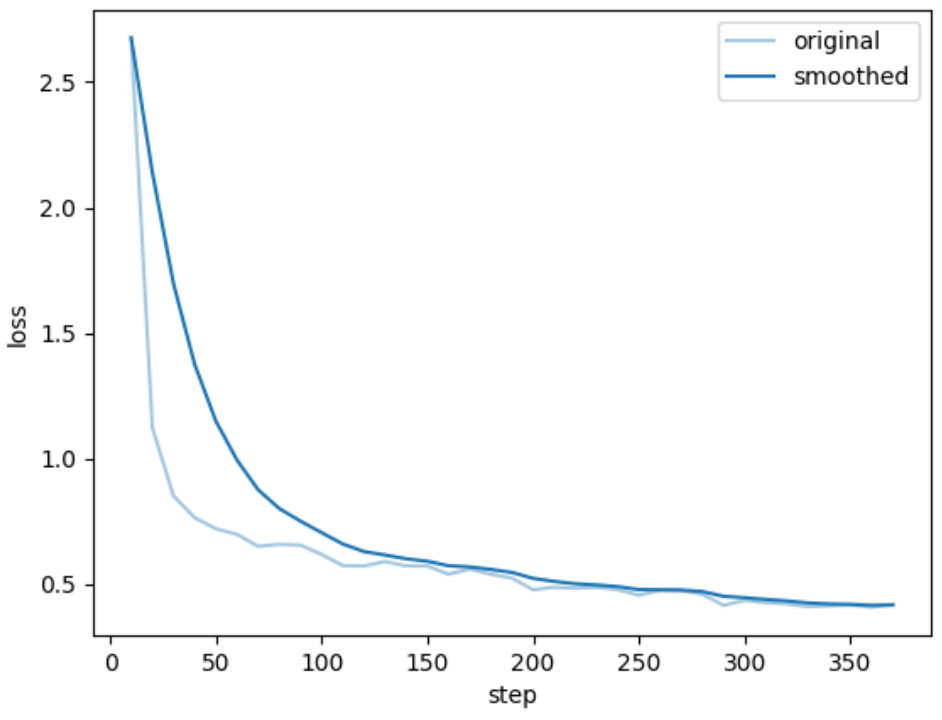

Training loss

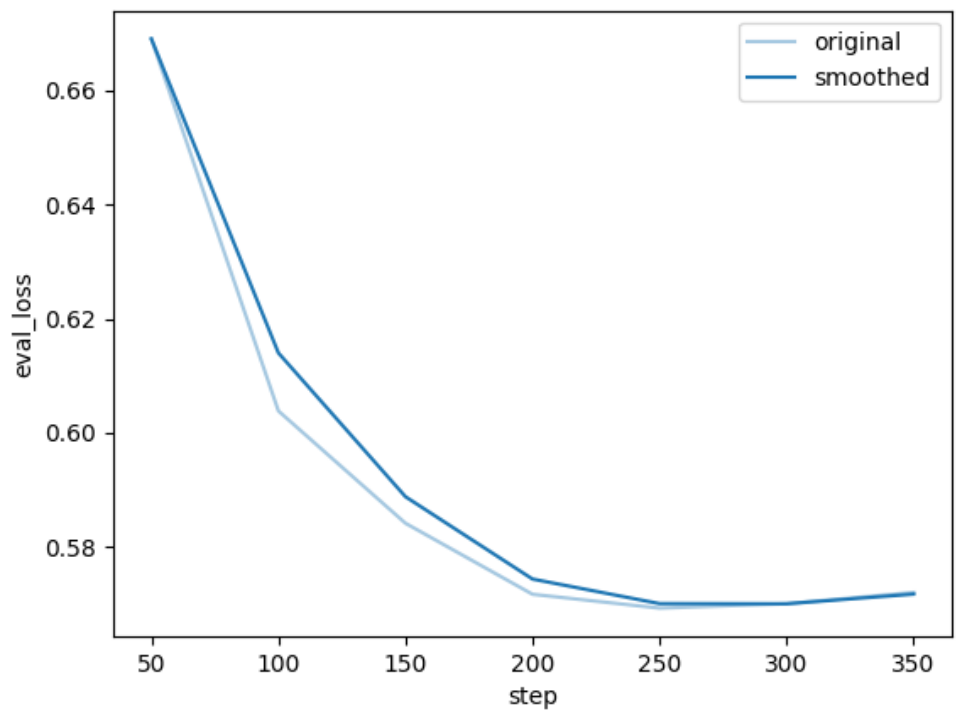

Evaluation loss

**Figure S3. Loss curves — without augmentation, with extra collection.**
Training loss (left) and evaluation loss (right) across training steps for all four cases.

**Canonical output schema, original training set**

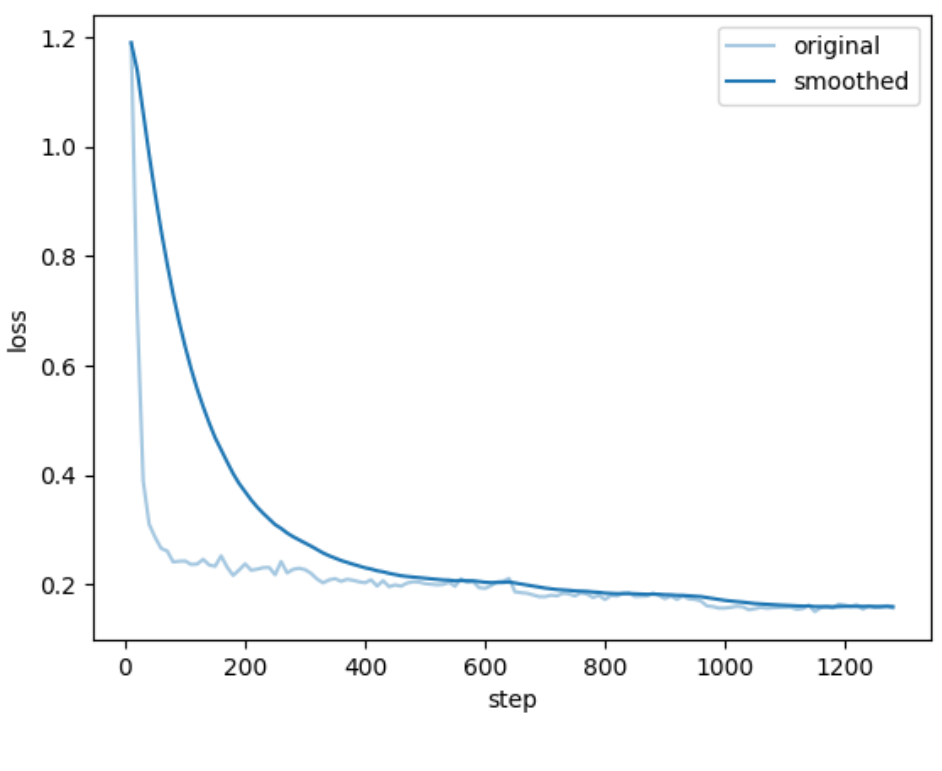

Training loss

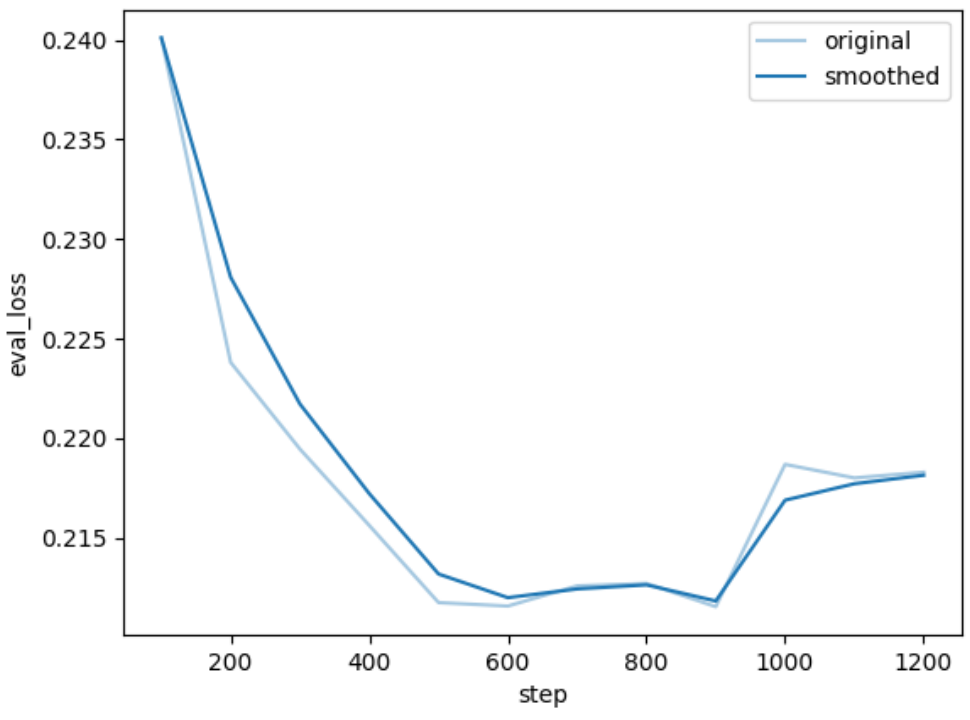

Evaluation loss

**Canonical output schema, rebalanced training set**

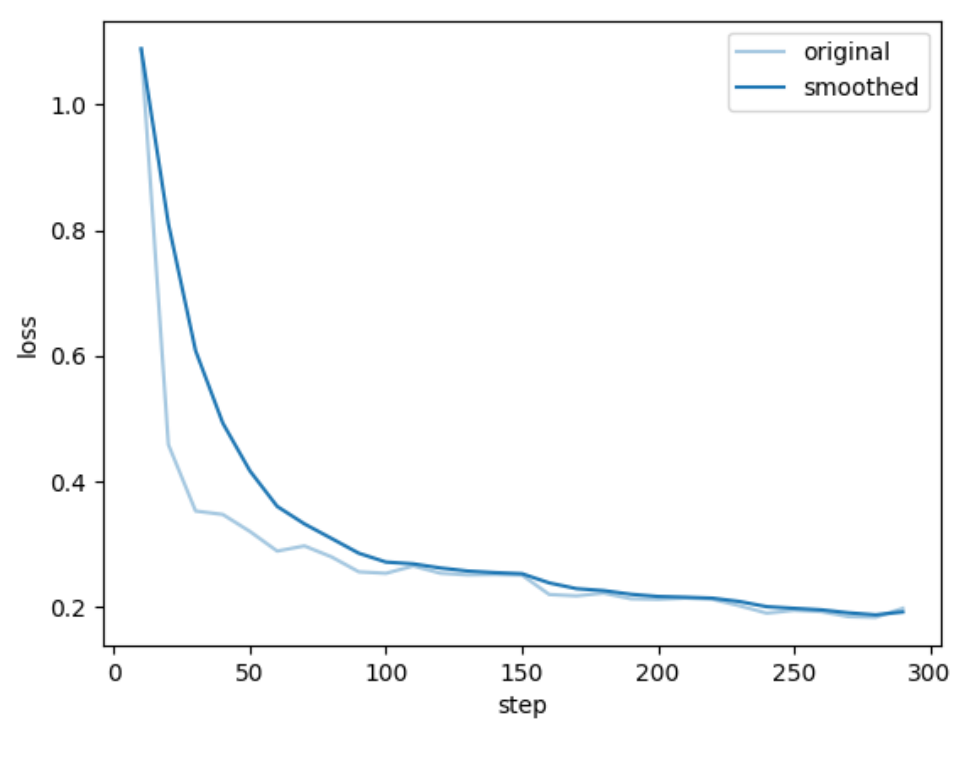

Training loss

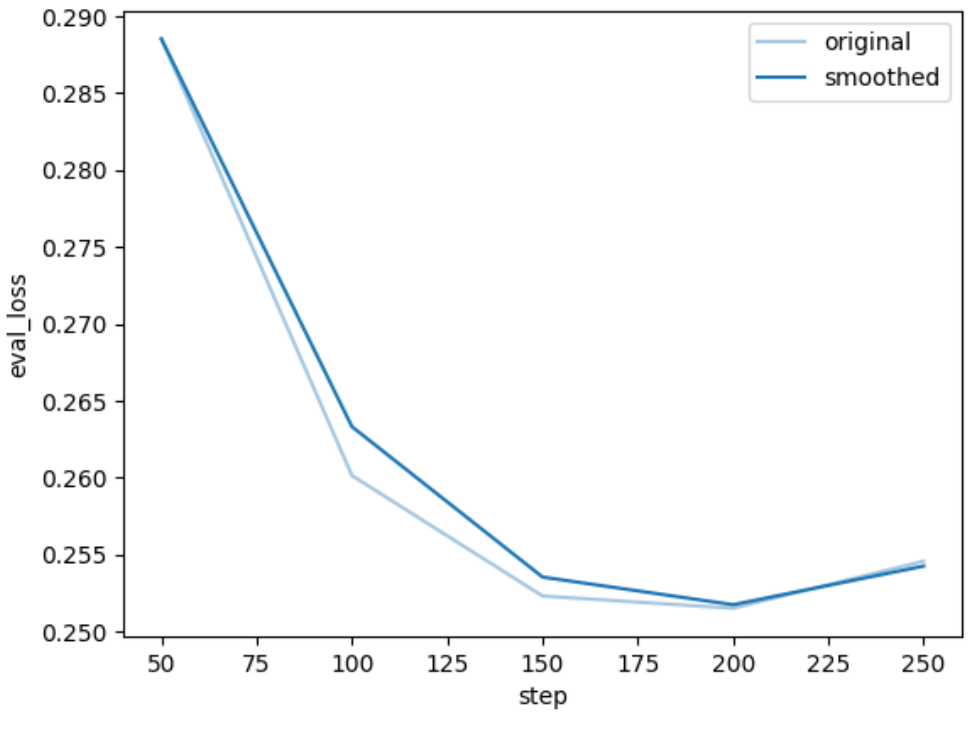

Evaluation loss

**Non-canonical output schema, original training set**

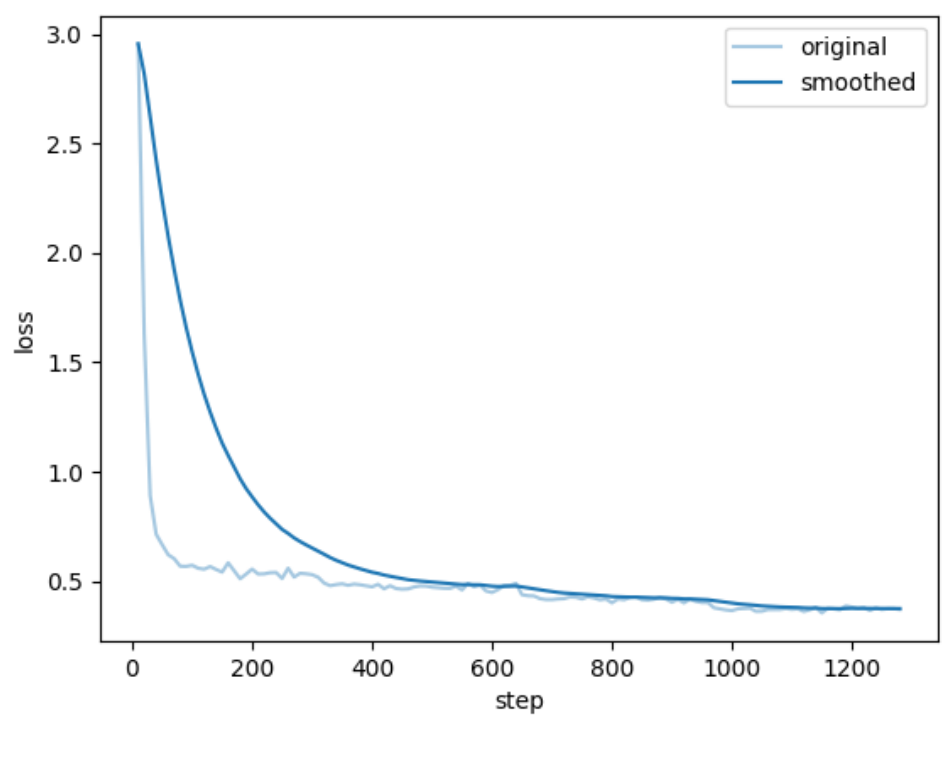

Training loss

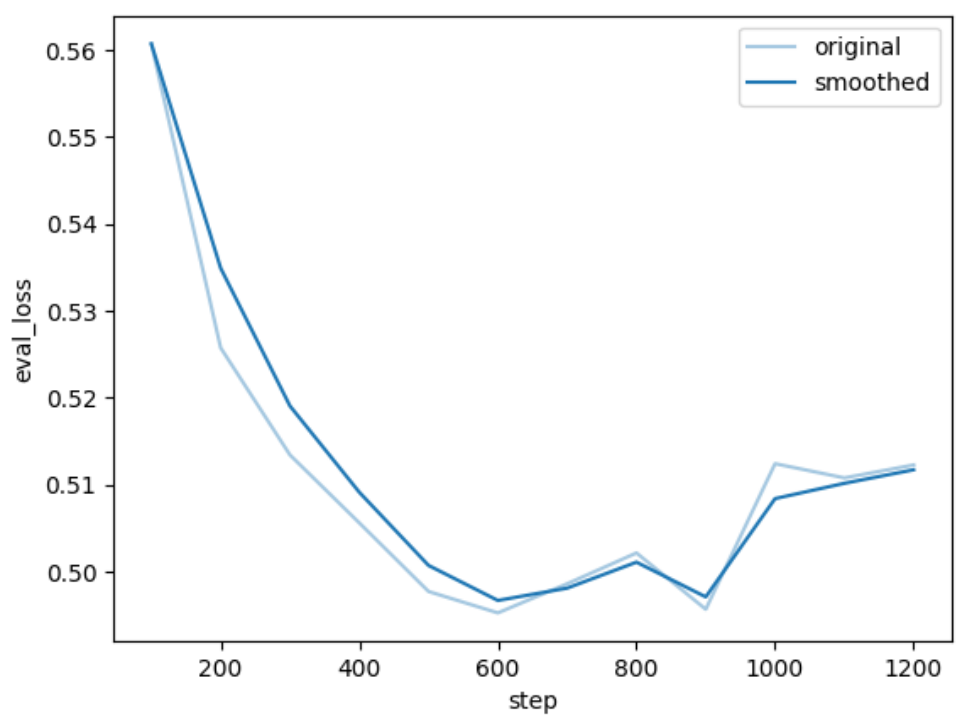

Evaluation loss

**Non-canonical output schema, rebalanced training set**

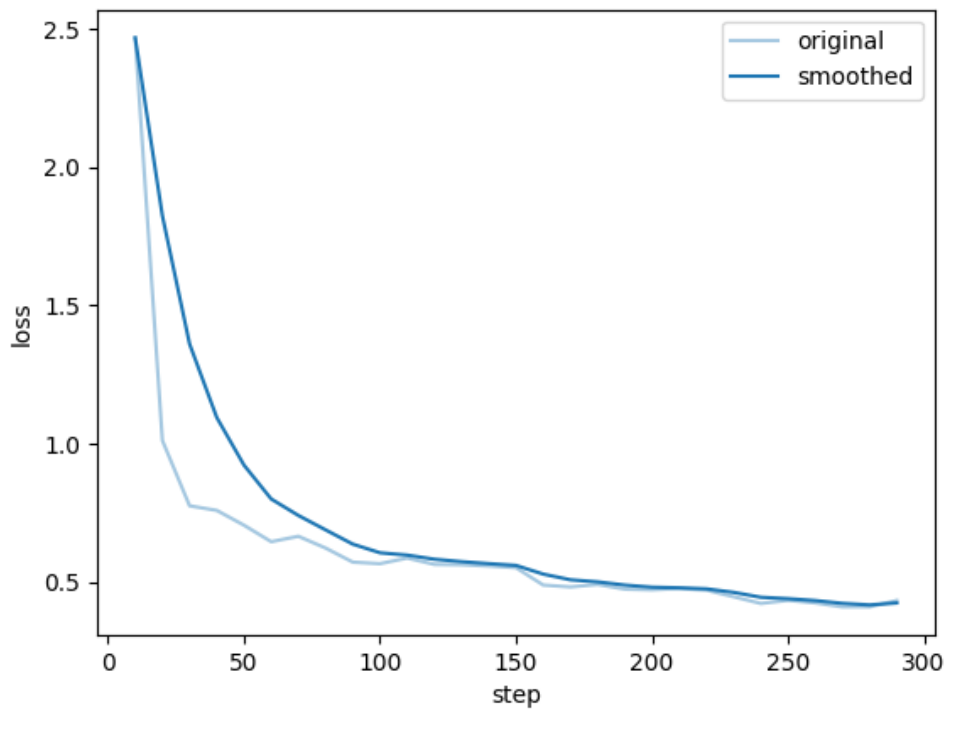

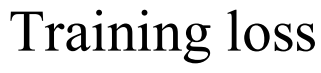

Training loss

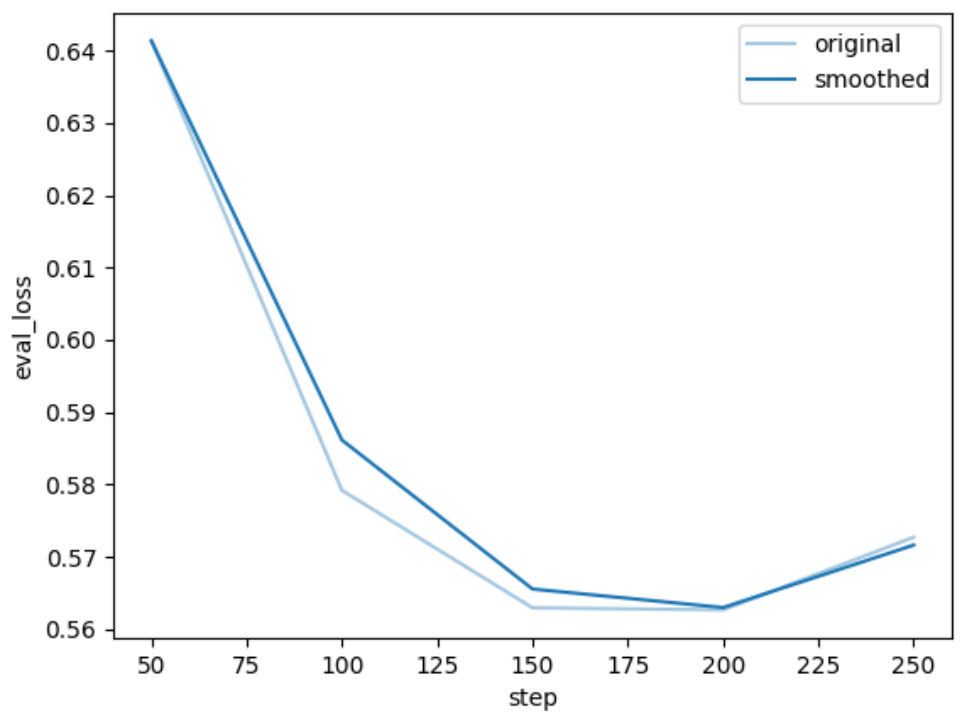

Evaluation loss

## Figure S4. Loss curves — without augmentation, without extra collection

Training loss (left) and evaluation loss (right) across training steps for all four cases.

### Canonical output schema, original training set

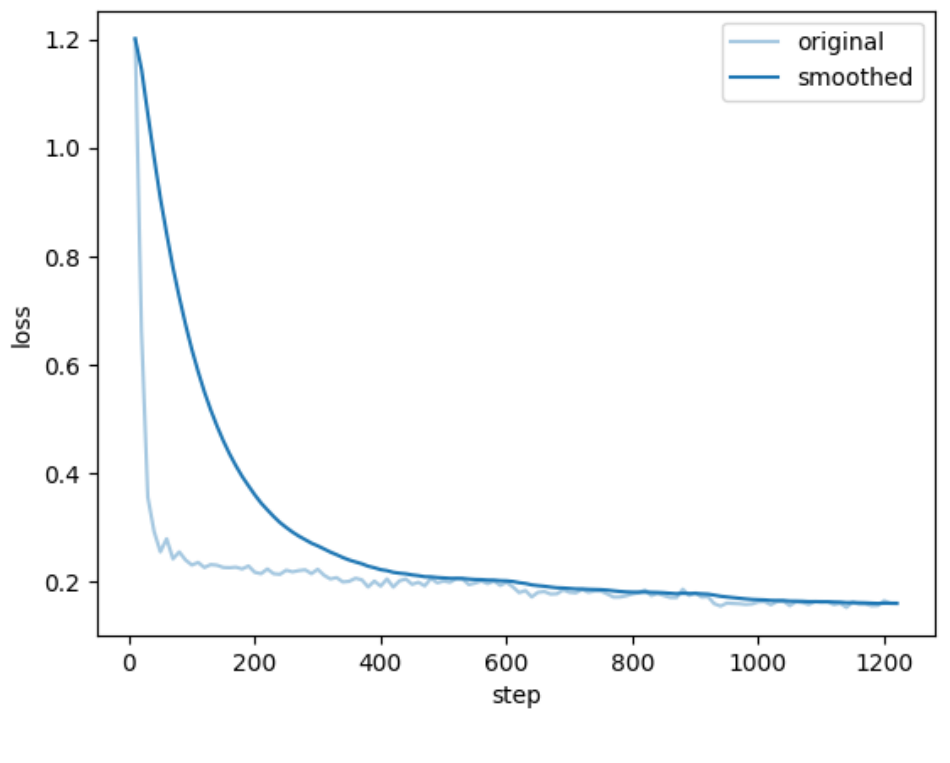

Training loss

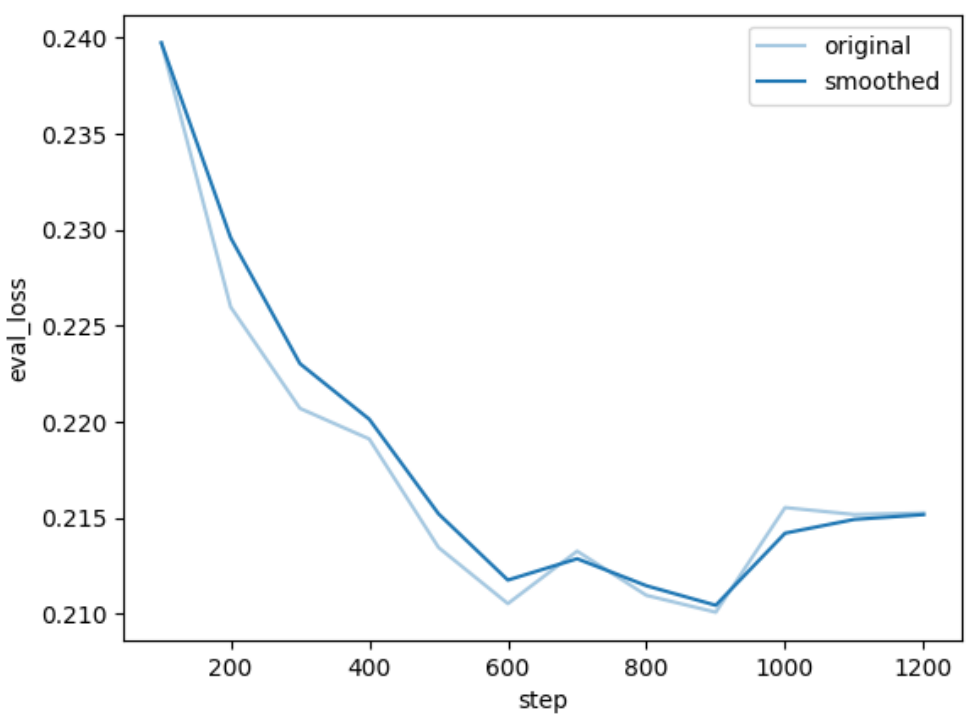

Evaluation loss

### Canonical output schema, rebalanced training set

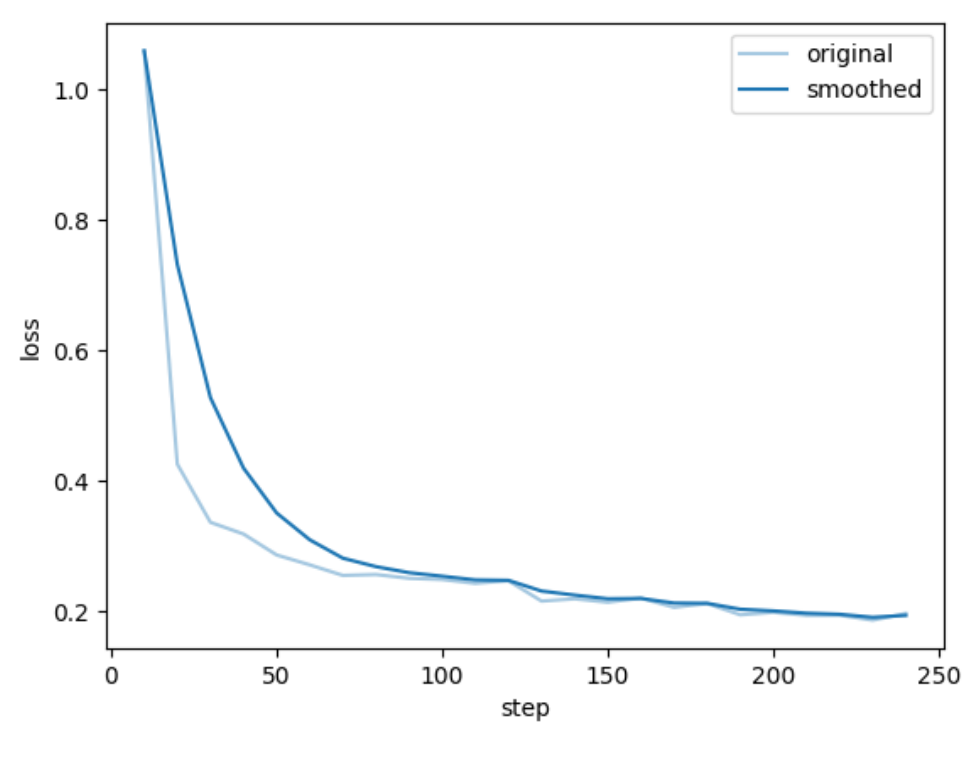

Training loss

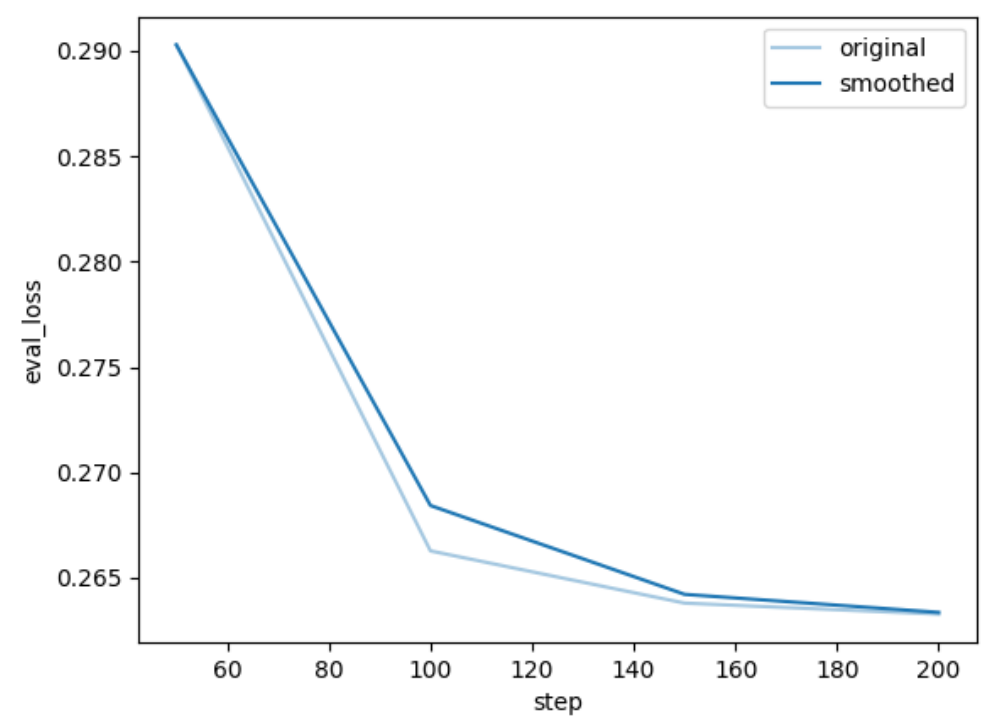

Evaluation loss

### Non-canonical output schema, original training set

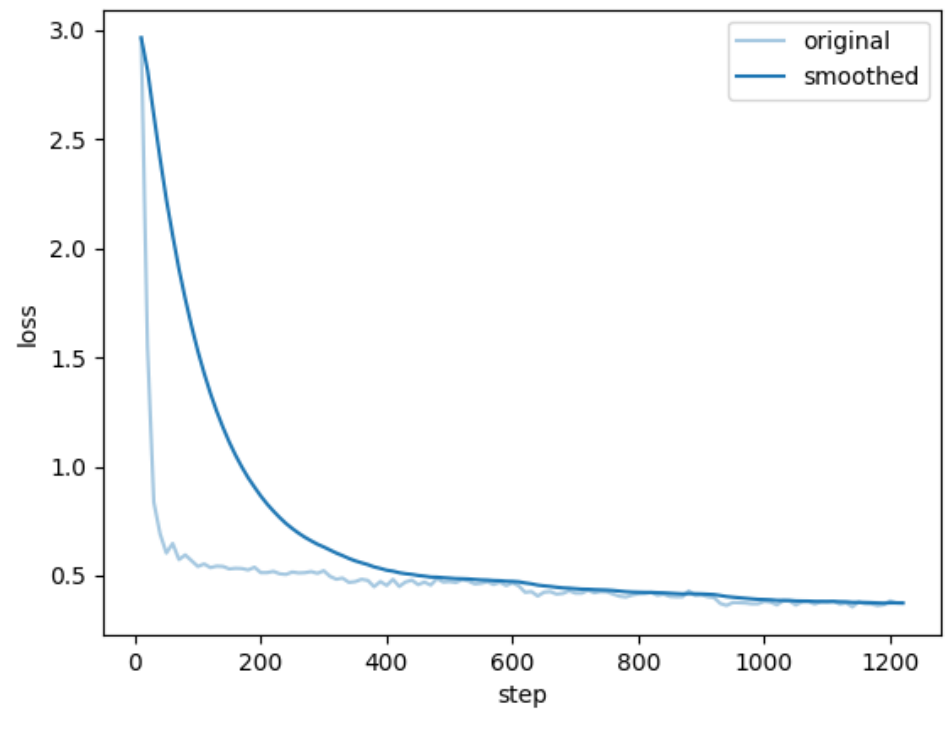

Training loss

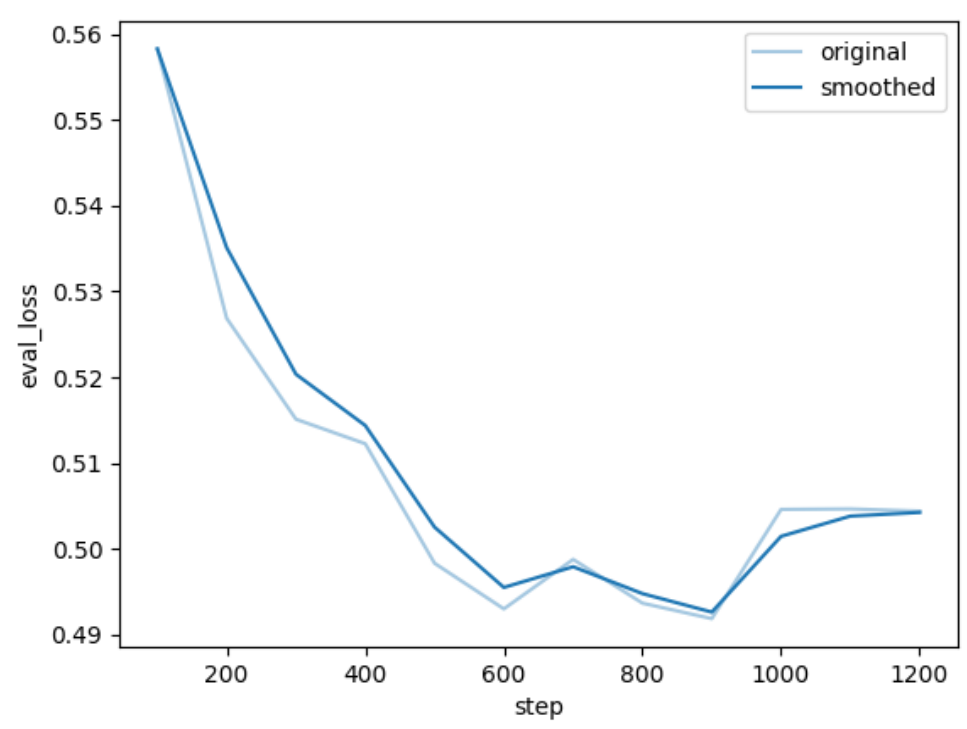

Evaluation loss

**Non-canonical output schema, rebalanced training set**

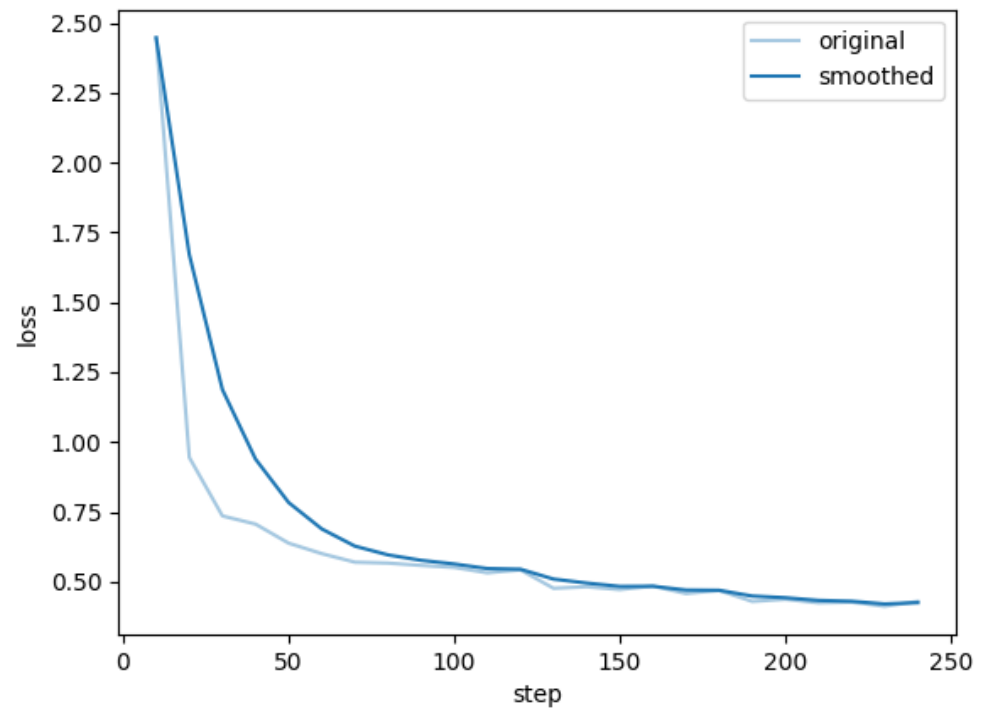

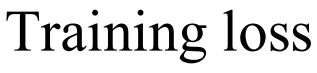
Training loss

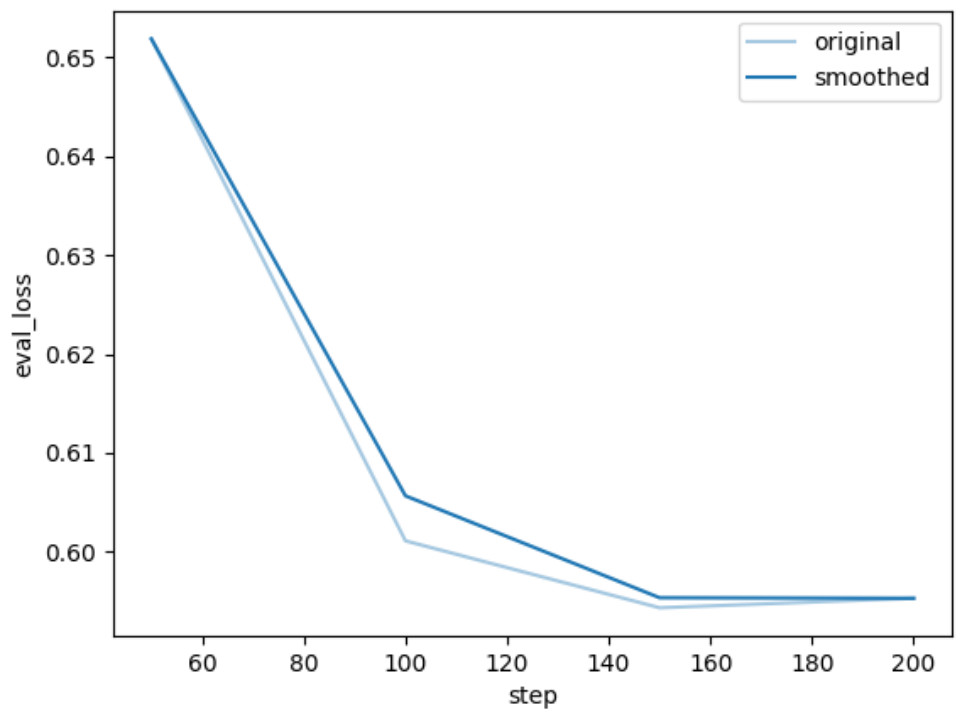

Evaluation loss

**Figure S5. Dataset-ablation results on the hold-out test set.**

Counterpart of **Fig. 5** of the main text, evaluated on the hold-out test set instead of the manual test set. Per-cell value reductions in image-averaged precision (left) and recall (right) for three fine-tuning variants (w/ augmentation & w/o extra collection; w/o augmentation & w/ extra collection; w/o augmentation & w/o extra collection), each evaluated under the four combinations of dataset split (Original, Rebalanced) and output schema (Canonical, Non-canonical). All value reductions are relative to the fully enriched reference configuration (w/ augmentation & w/ extra collection) on the hold-out test set; more negative (deeper red) entries indicate larger performance loss when that ingredient is removed.

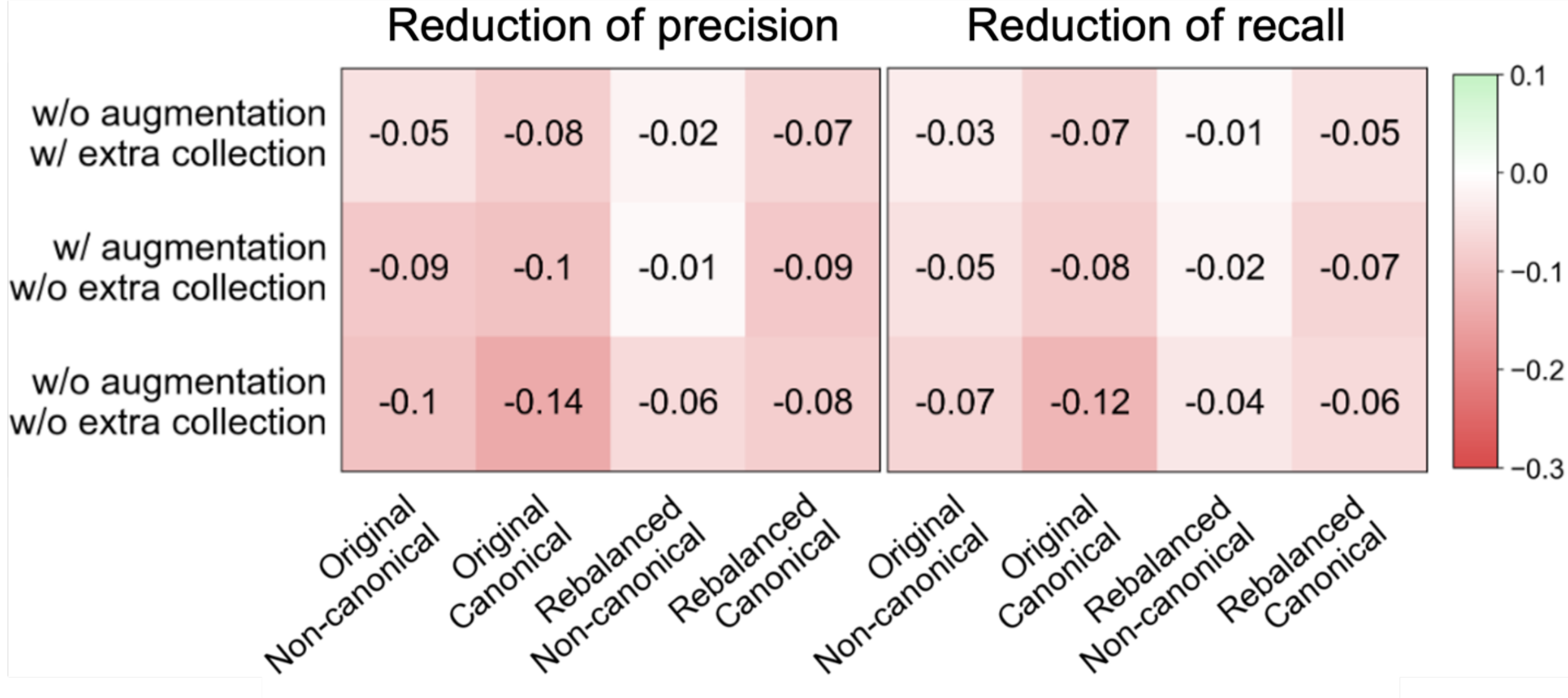

**Figure S6. Hold-Out test set results for proprietary models with specialist assistance.**
Counterpart of **Fig. 6** of the main text, evaluated on the hold-out test set instead of the manual test set. Per-cell performance gains in image-averaged precision (left) and recall (right) for GPT-5.5-Reasoning + FT-Qwen and Gemini 3.1 Pro-Reasoning + FT-Qwen relative to their unassisted baselines, each evaluated under the two output schemas (canonical, non-canonical). More positive (deeper green) entries indicate larger improvements from specialist assistance. Both proprietary reasoning models were run with reasoning effort set to high.

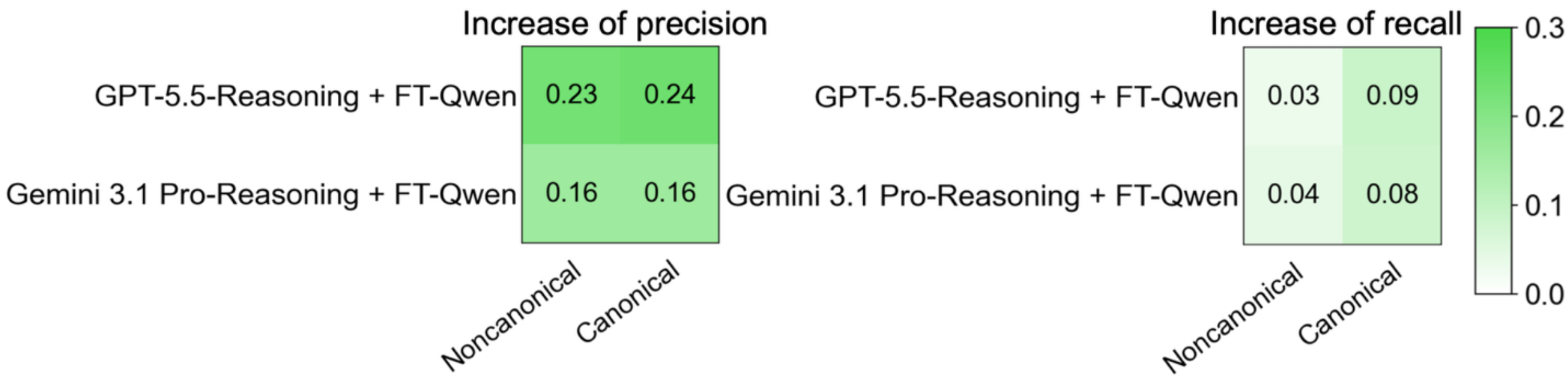

**Figure S7. Proprietary–specialist integration in fatigue fracture analysis: decoding-sweep judgment.**

Manual judgment of GPT-5.5-Reasoning (high) outputs for the case study in **Fig. 8**, comparing the unassisted baseline with the 10 specialist-assisted runs in **Table S3**. For Image 1, a response was judged correct only if it mentioned striation. For the second-condition views (Image 2, Image 2-1, and Image 2-2), a response was judged correct if it identified both striation on the facet and intergranular facet, partially correct if it identified only intergranular facet, and incorrect otherwise. Green, yellow, and red indicate correct, partially correct, and incorrect, respectively. The unassisted baseline was incorrect for Image 1 and partially correct for the second-condition views, whereas all specialist-assisted runs were correct.

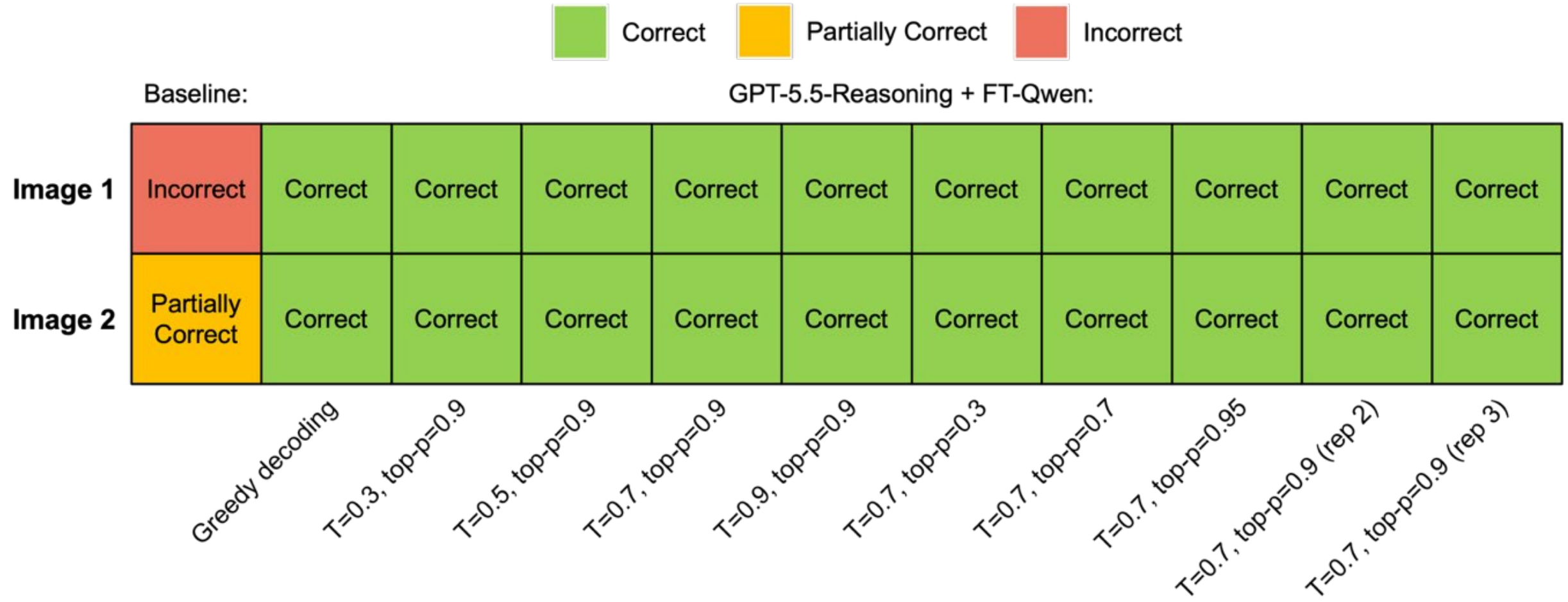